\documentclass[aps,prl, onecolumn,amsmath,amssymb,longbibliography,superscriptaddress]{revtex4-2}

\usepackage{amssymb,amsfonts,bm}
\usepackage{graphicx}
\usepackage{xcolor}
\usepackage{graphicx}
\usepackage{dcolumn}
\usepackage{ytableau}
\usepackage{epstopdf}
\usepackage{epsfig}
\usepackage[caption=false]{subfig}
\usepackage{hyperref}
\usepackage{verbatim}
\usepackage[export]{adjustbox}
\usepackage[english]{babel}
\usepackage{hyphenat}
\usepackage{scalerel}
\usepackage{url}
\usepackage{braket}
\def\aaa{\textbf{a}}
\def\bbb{\textbf{b}}
\usepackage{amsthm}
\usepackage{enumitem}
\usepackage{dsfont}
\def\tr{\operatorname{tr}}
\newcommand{\id}{\mathds{1}}
\newcommand{\al}{\alpha}
\newcommand{\dens}{f}
\newcommand{\be}{\begin{equation}}
\newcommand{\ee}{\end{equation}}
\newcommand{\bea}{\begin{eqnarray}}
\newcommand{\eea}{\end{eqnarray}}
\graphicspath{{numerics_plots}}
\def\tw{\tilde w}
\def\replN{N}
\def\hilbn{n}

\begin{document}

\title{A Dyson Brownian Motion Model for Weak Measurements in Chaotic Quantum Systems}

\author{Federico Gerbino}
\email{federico.gerbino@universite-paris-saclay.fr}
\affiliation{Laboratoire de Physique Th\'eorique et Modèles Statistiques, Université Paris-Saclay, CNRS, 91405 Orsay, France}

\author{Pierre Le Doussal}
\affiliation{Laboratoire de Physique de l'\'Ecole Normale Sup\'erieure, CNRS, ENS $\&$ PSL University, Sorbonne Universit\'e, Universit\'e Paris Cité, 
 75005 Paris, France}

\author{Guido Giachetti}
\affiliation{Laboratoire de Physique Th\'eorique et Mod\'elisation, CY Cergy Paris Universit\'e, 
CNRS, 95302 Cergy-Pontoise, France}

\author{Andrea De Luca}
\affiliation{Laboratoire de Physique Th\'eorique et Mod\'elisation, CY Cergy Paris Universit\'e, 
CNRS, 95302 Cergy-Pontoise, France}    

\begin{abstract}
We consider a toy model for the study of monitored dynamics in many-body quantum systems. We study the stochastic Schr\"odinger equation resulting from continuous monitoring with a rate $\Gamma$ of a random Hermitian operator, drawn from the Gaussian unitary ensemble (GUE) at every time $t$.
Due to invariance by unitary transformations, the dynamics of the eigenvalues $\{\lambda_\alpha\}_{\alpha=1}^n$ of the density matrix decouples from that of the eigenvectors, and is exactly described by stochastic equations that we derive.
We consider two regimes: in the presence of an extra dephasing term, which can be generated by imperfect quantum measurements, the density matrix has a stationary distribution, and we show that in the limit of large size $n\to\infty$ it matches with the inverse-Marchenko--Pastur distribution.
In the case of perfect measurements, instead,
purification eventually occurs and we focus on finite-time dynamics. In this case, remarkably, we find an exact solution for the joint probability distribution of $\lambda$'s at each time $t$ and for each size $n$. Two relevant regimes emerge: at short times $t\Gamma= O(1)$, the spectrum is in a Coulomb gas regime, with a well-defined continuous spectral distribution in the $n\to\infty$ limit. In that case, all moments of the density matrix become self-averaging and it is possible to exactly characterize the entanglement spectrum. In the limit of large times $t \Gamma = O(n)$, one enters instead a regime in which the eigenvalues are exponentially separated $\log(\lambda_\alpha/\lambda_\beta) = O(\Gamma t/n)$, but fluctuations $\sim O(\sqrt{\Gamma t/n})$ play an essential role.  We are still able to characterize the asymptotic behaviors of the entanglement entropy in this regime.
\end{abstract}

\maketitle

\section{Introduction}
In recent years, the study of quantum many-body systems in the presence of continuous monitoring of its local degrees of freedom has received much attention. The motivations for this case are of various kinds: first, from a practical point of view, the combinations of measurements and quantum gates are essential ingredients of quantum computation~\cite{PhysRevLett.101.010501,leung2004quantum}. In addition to this fact, recent studies have shown that quantum measurements (both projective and weak) can induce peculiar phase transitions, denominated measurement-induced phase transitions (MIPTs)~\cite{PhysRevX.9.031009, PhysRevB.98.205136,PhysRevB.100.134306}, which arise exclusively by looking at the statistics of trajectories via nonlinear observables~\cite{PhysRevX.13.011043,PhysRevX.13.011045,Bernard_2021,PhysRevB.99.174205}. Prime examples are transitions in entanglement dynamics~\cite{PhysRevX.9.031009, PhysRevB.98.205136,PhysRevB.100.134306} or purification time~\cite{PhysRevLett.125.070606} induced by increasing the strength of the measurements (e.g., the probability with which each site is measured in the case of projective measurements or the rate of measurements in the case of weak measurements). In this case, it was found that going from weak to strong measurements, one can move from a volume-law entangled phase~\cite{dalessio2016,PhysRevX.7.031016, PhysRevLett.111.127205} to an area-law one~\cite{PhysRevLett.109.017202,RevModPhys.91.021001,Calabrese_2005,PhysRevX.8.041019,PhysRevX.9.021033}. Results along these lines were initially obtained by using random circuits~\cite{doi:10.1146/annurev-conmatphys-031720-030658}, in which quantum evolution occurs through unitary gates chosen from a uniform distribution (Haar) possibly within an appropriate subset of the unitary group. Then, in the limit of a large local Hilbert space dimension, it has been shown that a phase transition is present, and the corresponding critical point lies in the directed percolation universality class. 
However, it has been clarified that this simplification is not accurate in general~\cite{PhysRevB.104.155111}: the critical point is described by a peculiar conformal (scale invariant) theory not exactly solvable in general~\cite{PhysRevB.101.104302,PhysRevLett.128.050602,nahum2023renormalization}. Following these results, much theoretical~\cite{PhysRevLett.128.010603,10.21468/SciPostPhysCore.6.1.023} and numerical~\cite{PhysRevResearch.2.013022,PhysRevB.106.024305,PhysRevLett.130.220404,PhysRevB.105.104306} work has been devoted to thoroughly characterizing MIPTs by linking them to various other phenomena, such as quantum error correction~\cite{PhysRevLett.125.030505}, purification~\cite{PhysRevX.10.041020,Ticozzi2014,loio2023purification}, state preparation, quantum communication \cite{kelly2023coherence}, and the complexity of classical simulation of quantum systems~\cite{PhysRevLett.91.147902,PhysRevLett.93.040502}. Recent results have attempted to study this phenomenology in simplified toy models~\cite{PhysRevResearch.4.043212, PhysRevResearch.2.023288}. However, in the case of noninteracting systems (e.g., free fermions), it has been shown that the transition has a radically different nature: the volume-law phase is immediately unstable for arbitrarily weak quantum measurements~\cite{10.21468/SciPostPhys.7.2.024,Fidkowski2021howdynamicalquantum,PhysRevB.105.094303,Santini2023Observation,PhysRevB.105.064305,PhysRevB.105.L241114,PhysRevB.103.224210}, and yet, in some cases, a subvolume phase can survive before the onset of the area-law phase~\cite{PhysRevLett.126.170602,PhysRevX.11.041004,PhysRevLett.128.010605,PhysRevResearch.4.033001,PhysRevA.107.032215}, with a universality class more akin to problems of Anderson localization and disordered conductors, studied in the context of nonlinear sigma models~\cite{fava2023nonlinear,poboiko2023theory}. 
In the interacting case, one of the main difficulties in studying individual trajectories of monitored systems lies in the fact that the probability of each trajectory depends on the state itself, in accordance with Born's rule. This aspect produces an inevitable nonlinearity when considering statistical averages over measurement outcomes, similar to that faced in disordered systems. In analogy to that approach, it is possible to proceed through the replica trick, thus studying $N$ identical copies of the system and eventually considering the relevant $\replN\to1$ limit (in contrast with the more common $\replN\to0$ limit relevant in disordered systems~\cite{mezard1987spin,A_J_Bray_1980}). 
The fundamental ingredient for the replica limit is the possibility of performing an analytic continuation in $N$ by means of an exact formula. Some studies avoid this problem by replacing the $\replN\to1$ limit with the simpler and more explicit $\replN=2$ case, where in fact two copies are considered. Alternatively, recent results have been obtained using mean-field models~\cite{PhysRevB.102.064202,PhysRevB.100.134203,bentsen2021measurement,PhysRevLett.127.140601}, where analytical progress can be achieved. In particular, in~\cite{PRXQuantum.2.010352}, a discrete fully connected model was introduced in which the dynamics in the presence of measurements can be mapped onto a variant of branching Brownian motion and then studied in terms of the {Fisher and
 Kolmogorov--Petrovsky--Piskunov reaction-diffusion Equation \cite{fisher1937wave, KPP, Derrida1988}}. However, the nature of these results in more general problems is not completely clear. More recently, a model of spin in the presence of noise and weak measurements has been introduced in which it is possible to study the dynamics in the presence of measurements analytically and in particular explicitly consider the $\replN\to1$ limit~\cite{giachetti2023elusive}: this showed that the replica limit can be  highly nontrivial, with a MIPT present for any integer $\replN > 1$, but disappearing in the $\replN\to1$ limit. 

Beyond the study of the transition, it becomes interesting to characterize the specific behavior of the two phases. In particular, the volume-law phase is hard to simulate from the classical point of view because of the strong presence of entanglement surviving very long times. In this paper, we focus on the characterization of the dynamics in the presence of weak measurements far from the critical point, within the volume-law phase. To do so, we introduce a model of the evolution of the density matrix of the system purely based on random matrices, akin to the one considered in Ref.~\cite{schomerus2022noisy}. We choose a dynamics based on the stochastic Schr\"odinger Equation \cite{PhysRevA.36.5543}, invariant under unitary rotations in the $n$-dimensional Hilbert space, such that the dynamics of the eigenvalues of the density matrix decouples from that of the eigenvectors, as is the case in Dyson Brownian motion (DBM). In this way, unitary dynamics becomes inessential since it does not affect the eigenvalues.
Once we have derived a stochastic equation for the evolution of the eigenvalues, we study two relevant limits of it: in the first case, we consider dynamics induced by ``imperfect'' measurements, i.e., a case in which a fraction of the measurement results is unknown. This is effectively equivalent to introducing a dephasing term that on each trajectory tends to bring the density matrix closer to the identity. In this case, we prove the existence of a steady state that describes the distribution of the eigenvalues: at large $n$, the resulting ensemble is described by an inverse-Wishart matrix.
Next, we explore the dynamics for perfect measures: in this case, the steady state at long times reduces trivially to a pure random state uniformly distributed in the steady state; however, this happens with a characteristic transient at small and intermediate times. We present the exact solution for the joint distribution of eigenvalues at all times and analyze its effects on the entanglement~entropy.

\section{Preliminaries on Unravelings and Trajectories}
{The combined dynamics of a quantum system undergoing both unitary evolution and measurements can always be modeled as a quantum channel~\cite{nielsen2002quantum}, whose output contains both the state of the quantum system and the classical encoding of the results of the measurements.
According to Choi's theorem~\cite{CHOI1975285}, a quantum channel $\Phi(\rho)$ can always be expressed in terms of a set of Kraus operators $\{\mathcal{K}_{\aaa}\}_{\aaa}$, first discussed in Ref.~\cite{KRAUS1971311}. 
Those provide an explicit operator-sum decomposition of the quantum channel as $\Phi(\rho) = \sum_\aaa \mathcal{K}_{\aaa} \rho \mathcal{K}_{\aaa}^\dag$, and describe the dynamics by means of linear maps from the system's Hilbert space to itself. 
Because of trace-norm conservation, Kraus operators satisfy the condition $\sum_\aaa \mathcal{K}_{\aaa}^\dag \mathcal{K}_{\aaa} = \id$, with $\id$ the identity operator in the Hilbert space of the quantum system. }
However, the specific decomposition of $\Phi$ in terms of the Kraus operators is not unique as $\{\mathcal{K}_{\aaa}'\}_{\aaa}$ with $\mathcal{K}_{\aaa}' = \sum_\bbb U_{\aaa,\bbb}\mathcal{K}_{\bbb}$, for an arbitrary unitary transformation $U_{\aaa, \bbb}$, define the same quantum channel $\Phi(\rho)$. Different decompositions of the same quantum channel are referred to as unravelings.
In the context of repeated measurement, 
the Kraus operators are factorized $\mathcal{K}_{\aaa} = K_{a_1} K_{a_2}\ldots K_{a_T}$ where  $\aaa = (a_1,a_2,\ldots, a_T)$ labels the collection of all measurement outcomes performed at each time step. For MIPTs, one is interested in the single trajectory where the initial density matrix $\rho(0)$ evolves as
\begin{equation}
\label{eq:rhoa}
    \rho_\aaa = p^{-1}_\aaa \tilde\rho_\aaa \;, \quad \tilde \rho_\aaa \equiv \mathcal{K}_{\aaa} \rho(0) \mathcal{K}_{\aaa}^\dag \,,
\end{equation}
where the probability of a specific trajectory has been introduced according to Born's rule as $p_\aaa = \tr[\tilde \rho_\aaa]$. Given any functional of the state $F[\rho]$, we define the average over trajectories~as
\begin{equation}
\label{eq:BR}
    \langle F[\rho] \rangle = \sum_\aaa p_\aaa F[\rho_\aaa] \,, \quad \mbox{Born rule.}
\end{equation}
If the results of the measurements are not known, one only has access to linear functionals of the state, such as the quantum expectation of any observable, e.g., $\langle \tr[\hat O \rho] \rangle = \tr[\hat O \Phi[\rho(0)]]$, which depends solely on the quantum channel $\Phi[\rho]$ and is independent on the specific unraveling. However, this is not true for nonlinear functionals such as the Renyi's entropies, where one considers the functionals
\begin{equation}
\label{eq:renyidef}
S_k [\rho] := \frac{1}{1-k} \ln \tr[\rho^k] \,,
\end{equation}
and in particular, the Von Neumann entropy defined as the limit $k\to 1$, or, explicitly,
\begin{equation}
    \label{eq:VN_entropy_def}
S_1 [\rho]  := -\tr[\rho \log \rho] = -\sum_\alpha \lambda_\alpha \log \lambda_\alpha \,,
\end{equation}
where $\{\lambda_\alpha\}_{\alpha=1}^\hilbn$ denotes the eigenvalues of $\rho$. Quantities like $\langle S_1[\rho]\rangle $ can be used as order parameters for MIPTs.
By discarding the Born's weight for trajectories, one can define a different average
\begin{equation}
\label{eq:forcemeas}
    \langle F[\rho] \rangle_0 = \frac{\sum_\aaa F[\rho_\aaa]}{\sum_\aaa 1} \,, \quad \mbox{Unbiased outcomes}\,,
\end{equation}
where all measurement outcomes have equal weight and are statistically independent at different times. Unbiased outcomes can be obtained via post-selection by artificially attributing the same probability to each trajectory. In some contexts, $\langle \ldots \rangle_0$ is used as an approximation for the more physical $\langle \ldots \rangle$. We will see that both cases emerge naturally in our framework.

\section{The model}
We consider a toy model for continuous monitoring in a quantum system with a Hilbert space of dimension $\hilbn$. Let us explain the model by splitting the unitary evolution and the monitoring parts.
For the unitary evolution, we have the Hamiltonian increment
\begin{equation}
    dH = J \sum_{\alpha, \beta=1}^\hilbn dh_{\alpha\beta} (t) \ket{\alpha}\bra{\beta} \,,
\end{equation}
where hermiticity is ensured requiring that $dh_{\alpha \beta} = dh_{\beta \alpha}^\ast$.
Otherwise, the increments are chosen as a Hermitian Brownian motion with covariance $dh_{\al\beta} dh_{\gamma\delta}^* =  \delta_{\al\gamma}\delta_{\beta\delta} \,dt / \hilbn$. 
Equivalently, in the limit $dt \to 0$, we can write $dh_{\alpha\beta} = \sqrt{dt}\, h_{\alpha\beta}$, where $h_{\alpha\beta}$ is an $\hilbn \times \hilbn$ Hermitian matrix, drawn from the Gaussian unitary ensemble (GUE)
\begin{equation}
\label{eq:GUE}
    P(h) =  \frac{1}{Z} e^{-\frac{\hilbn}{2} \tr[h h^\dag]}  \,,
\end{equation}
and the measure is over the independent entries of a Hermitian matrix (with a semi-circle spectrum of support $[-2,2]$ in the
large $n$ limit). A crucial property of the GUE distribution is its invariance under unitary conjugation, $P(h) = P(u h u^\dag)$ for any unitary $u \in U(n)$. 
The dynamics induced by such a unitary evolution is simply described by $\rho + d\rho = e^{- i dH} \rho e^{i dH}$, where the It\^o calculus conventions are implied.
For measurements, let us first introduce the standard form of the stochastic Schr\"odinger Equation \cite{PhysRevA.36.5543,PhysRevA.58.1699,Gisin_1992} for an imperfect monitoring of a Hermitian operator $\hat O$. Within a small time step $\Delta t$, the system is evolved coupling the operator $\hat O$ with an extra spin $1/2$ (ancilla) whose value along a reference direction is projectively measured with an outcome $a = \pm 1$. As explained in the previous section, this evolution can be encoded into the Kraus operators $K_a = (1 - a \sqrt{\Gamma \Delta t} \hat O + \gamma \Delta t \hat{O}^2 /2)/\sqrt{2}$ (see Appendix A in Ref.~\cite{giachetti2023elusive} for a summary of this derivation), where $\Gamma$ is a rate quantifying the measurement strength. In the limit $\Delta t \to 0$, the collection of measurements $\aaa$ can be used to define a standard Wiener process $Y(t)$ ($dY^2 = dt$) and one obtains the stochastic evolution equation
\begin{equation}
\label{eq:lindblbiasmeas0}
d\rho = (1 + x)\, \Gamma dt \, \mathcal{D}_O[\rho] + \sqrt{\Gamma} dY \{\hat O - \langle \hat O \rangle_t, \rho\} 
\end{equation}
with the dephasing superoperator $\mathcal{D}_O[\rho] \equiv  \hat O \rho \hat O - \frac 1 2 \{\hat O^2, \rho\} $. The parameter $x\geq 0$ quantifies an extra source of dephasing. It can be seen as due to a fraction of measurements whose outcomes are not known or more generally to a coupling with an external dephasing bath. The limit $x=0$ corresponds to perfect measurements. 
In our model, within each infinitesimal time step $dt$, we choose $\hat{O}(t) = \sum_{\alpha,\beta=1}^n o_{\alpha\beta}(t) \ket{\alpha}\bra{\beta}$ to be a random operator, with components $o_{\alpha\beta}$ drawn from the GUE distribution \eqref{eq:GUE}. We can thus introduce a Hermitian Brownian motion setting $do_{\al\beta} = o_{\al\beta}(t) dY$, satisfying $do_{\al\beta}do_{\gamma\delta}^* =  \delta_{\al\gamma}\delta_{\beta\delta} \, dt / \hilbn$. Using this, according to the rules of stochastic calculus, we can rewrite the dephasing part as $\mathcal{D}_O[\rho] dt = \mathcal{D}_O[\rho] dY^2 = d\hat{O} \rho d\hat{O} - \frac 1 2 \{ d\hat{O}^2, \rho \} = 
- (\rho - \frac{\id}{n})dt$, where $d\hat{O}= \sum_{\alpha,\beta=1}^n do_{\alpha\beta} \ket{\alpha}\bra{\beta}$. Note that at finite $\Delta t$, $o_{\alpha\beta}(t) dY$ is the product of two Gaussian distributions that is not Gaussian. However, this is irrelevant in the $\Delta t \to 0$ limit, where only the covariance is relevant to define a Wiener process.

In the following, we will assume that the density matrix is initially prepared in the infinite-temperature state $\rho = \id/n$. Thus, because both the Hamiltonian increment and the observable $\hat O$ are always chosen from the GUE, the distribution of the density matrix at any time is itself invariant under unitary transformation. In other words, diagonalizing $\rho = u \Lambda u^\dag$, $u$ will be Haar distributed in the unitary group $U(n)$ and will completely decouple from the dynamics of $\Lambda  = \operatorname{diag}(\lambda_1,\ldots, \lambda_n)$. In the following, we will focus on the dynamics and the distribution of the eigenvalues $\{\lambda_\al\}_{\al=1}^n$. As they are unaffected by the unitary dynamics, in the following we will ignore the latter.

\section{The \boldmath{$\hilbn = 2$} Case} \label{sec:n2}
Before dealing with the general $n$ case, it is worth considering specifically the $n=2$ case as an instructive warm-up.  We can parameterize the density matrix in terms of a vector $\vec r = \{r_1,r_2,r_3\}$, within the Bloch sphere ($|\vec{r}| \leq 1$), as 
\begin{equation}
    \rho = \frac 1 2 ( \id + \Vec{r} \cdot\Vec{\sigma}) = \frac{1}{2} (\id + r_\alpha \sigma_\alpha )\,,
\end{equation}
where $\vec{\sigma}=\{\sigma_1,\sigma_2,\sigma_3\}$ denotes the set of Pauli matrices. The modulus $r = |\vec r|$ coincides with the difference $|\lambda_1-\lambda_2|$ between the two eigenvalues of $\rho$, whereas the normalization of the trace, forcing $\lambda_1 + \lambda_2 = 1$, is implicit in the parameterization. 
It is also convenient to similarly parameterize the measurement operator as $O=o_0\id + \vec{o}\cdot\vec{\sigma}$. We can assume $\hat{O}$ to be traceless so that $o_0=(o_{11}+o_{22})/2=0$, as this contribution would be inessential in either case. The three remaining one-indexed random variables $o_1 = \mathrm{Re} (o_{12})$, $o_2=-\mathrm{Im}(o_{12})$, $o_3=(o_{11}-o_{22})/2$ are real-valued and satisfy $o_\alpha o_\beta = \delta_{\alpha \beta} /4$. In terms of these variables, the infinitesimal variation $dr_\alpha= \tr (d\rho \sigma_\alpha)$ reads
\begin{equation}
    dr_\alpha = - \Gamma dt (1+x) r_\al + 2\sqrt{\Gamma} (do_\alpha - do_\beta r_\beta r_\alpha) \,.
\end{equation} 
This can be recast into a closed stochastic equation for the modulus $r=|\vec{r}|$ (as required by unitary invariance), which takes the form \\
\begin{equation}
\label{eq:r}
    dr = \Gamma dt \left( \frac{1-r^2}{r} - x r \right) + \sqrt{\Gamma} dY (1-r^2)\,,
\end{equation} 
$dY$ being the standard Wiener process satisfying $dY^2=dt$ defined before Equation~\eqref{eq:lindblbiasmeas0}. In order to recast the previous equation into the standard Langevin form, let us perform the change in variables $r = \mathrm{tanh}(\omega)$, $\omega \in \mathbb{R}^{+}$. In terms of $\omega$, we indeed have $d\omega = - \Gamma V'(\omega) dt + \sqrt{\Gamma}dY$, where the potential is given by the following: 
\begin{equation} \label{eq:V2}
    V(\omega)= - \frac x 2  \mathrm{cosh}^2(\omega) - \log[\mathrm{sinh}(2\omega)] \,. 
\end{equation}
As a consequence, the evolution of the probability distribution $P(\omega,t)$ can be described through the associated Fokker--Planck (FP) equation
\begin{equation} \label{eq:FP2}
    \partial_t P(\omega,t) =  \frac{\Gamma}{2} \partial_\omega \left[ \left(\partial_\omega + 2 V'(\omega) \right) P(\omega,t) \right] \,.
\end{equation}
which admits a stationary probability distribution $P_{\mathrm{stat}}(\omega) = \frac{1}{Z} e^{-2V(\omega)} $. Coming back to the original coordinate, one has
\begin{equation} \label{eq:stat_distrib_n2}
    P_{\mathrm{stat}}(r) = \frac 1 Z \frac{r^2}{(1-r^2)^3} e^{-x(1-r^2)^{-1}}\,,
\end{equation}
a result already derived in~\cite{giachetti2023elusive} in the context of mean-field approximation. Since $\tr(\rho^2) = (1 + r^2)/2$, it follows that the purity has a nontrivial stationary distribution as well for any $x > 0$. As $x \rightarrow 0$, the distribution becomes more and more peaked around $r = 1$, eventually collapsing to $P_{\mathrm{stat}}(r) = \delta(r-1)$ for $x=0$: this is consistent with the fact that perfect measurements eventually lead to purification. For $x=0$, the finite-time solution of Equation~\eqref{eq:FP2} can be explicitly worked out as follows:
\begin{equation} \label{eq:distrib_n2}
    P(\omega, t) = \frac{\omega  \sinh (2 \omega ) e^{-\frac{\omega ^2}{2 \Gamma  t}-2 \Gamma  t}}{\sqrt{2\pi } (\Gamma  t)^{3/2}} \,,
\end{equation}
normalized in $\omega \in \mathbb{R}^+$.
The origin of this form will be clarified for general $n$ in Section~\ref{subsec:exact_finite_time}.

We can use these results to compute the Von Neumann entropy~\eqref{eq:VN_entropy_def}. In particular, on each trajectory, it can be related to the variable $r$
\begin{equation}
    S_1  = S_{\rm max} - \frac{1}{2}  \left[(1+r) \log(1+r)+ (1-r) \log(1-r) \right] \,,
\end{equation}
with $S_{\rm max}=\log n=\log 2$, the entropy of the maximally mixed state.
We can compute the behavior of $\langle S_1 \rangle$ in different regimes; in the stationary state at $x>0$, one has
\begin{equation}
\langle S_1 \rangle_{\rm stat} = 
\begin{cases}
\log 2 - \frac{3}{4x} + O(x^{-2})\,, & x\gg1 \\
- \frac 1 4 x\log x \,,& x\ll 1
\end{cases} \,.
\end{equation}
For $x=0$, the system purifies, and $\langle S_1 \rangle \rightarrow 0$. For large times, we have
\begin{equation} \label{eq:VN_t_n=2}
    \langle S_1 \rangle = \frac{e^{-2 \Gamma t}}{2} \left( 1 + \frac{(\Gamma t)^{-\frac{1}{2}}}{\sqrt{2 \pi}} - \frac{\pi^2+6}{24} \frac{(\Gamma t)^{-\frac{3}{2}}}{\sqrt{2 \pi}} +  O(t^{-\frac{5}{2}})  \right) \,.
\end{equation}
See Appendix~\ref{finitetimeS} for the details of the derivation and for the higher orders.

\section{Dynamics of the Spectrum}
\subsection{Stochastic Evolution of the Eigenvalues}
In order to derive an evolution equation for the eigenvalues in the general case, we follow the standard approach~\cite{10.1063/1.1703862} and make use of second-order perturbation theory, as follows: 
\begin{equation}
    d\lambda_\al = d\rho_{\al\al} + \sum_{\beta\neq \al}\frac{|d\rho_{\al\beta}|^2}{\lambda_\al-\lambda_\beta} \,.
\end{equation}
Now, one has to substitute the infinitesimal variations $\rho_{\al}$ as determined in Equation~\eqref{eq:lindblbiasmeas0}. To further simplify the calculation, we take advantage of unitary rotational invariance, so that $\rho$ at time $t$ is assumed to be diagonal in an appropriate basis $\rho_{\al\beta} = \lambda_\al \delta_{\al\beta}$ (no sum over $\alpha$). Thus, we have 
\begin{equation}
    d\rho_{\al\al} = -\Gamma dt \,(1+x) \left(\lambda_\al - \frac{1}{n}\right) + 2 \sqrt{\Gamma} \lambda_{\al} \big[ do_{\al\al} - \sum_\beta \lambda_\beta do_{\beta\beta} \big] \,,
\end{equation}
whereas for the off-diagonal elements, we simply have
$d\rho_{\al \neq \beta} = \sqrt{\Gamma}  do_{\al \beta} (\lambda_\al + \lambda_\beta )$. We can consider the diagonal elements of $d\hat{O}$ and rescale them as $dB_\alpha = \sqrt{n} \,do_{\al\al}$, so that $dB_\alpha dB_\beta = \delta_{\alpha\beta} \,dt$ are standard Wiener increments. Then, one finally finds 
\begin{equation} \label{eq:eigen}
    d \lambda_\al = \gamma dt \,\left[ - (1+x)(n\lambda_\al -1) + \sum_{\beta\neq\al} \frac{(\lambda_\al+\lambda_\beta)^2}{\lambda_\al-\lambda_\beta} \right] + 2\sqrt{\gamma} \lambda_\al \left(dB_\al - \sum_\beta \lambda_\beta dB_\beta \right) \,, 
\end{equation}
where we rescaled $\gamma = \Gamma / n$.
Let us notice that, as it should be, the evolution of Equation~\eqref{eq:eigen} preserves the trace, as $\sum_\al d\lambda_\al = d\tr(\rho) = 0$. Additionally, for $x=0$, the configuration $\lambda_\al=1$, $\lambda_{\beta\neq\al}=0$ is a fixed point, since as expected, in the absence of any dephasing ($x=0$), measurements eventually lead to purification and the density matrix reduces to a rank-$1$ projector.
It is useful to further manipulate Equation~\eqref{eq:eigen_x} by rewriting $\lambda_\al-1/n = \frac1 n \sum_{\beta\neq\al}(\lambda_\al - \lambda_\beta)$, where we used $\sum_\beta \lambda_\beta =1$. In such a way, we obtain
\begin{equation} \label{eq:eigen_x}
    d \lambda_\al = \gamma dt \sum_{\beta\neq\al} \left[ 4 \frac{\lambda_\al\lambda_\beta}{\lambda_\al-\lambda_\beta} + x (\lambda_\beta-\lambda_\al) \right] + 2\sqrt{\gamma} \lambda_\al \left(dB_\al - \sum_\beta \lambda_\beta dB_\beta \right) \,.
\end{equation}
For $x=0$, Equation~\eqref{eq:eigen_x} presents some analogy with the eigenvalue flow studied in \cite{gautie2021matrix}, see 
Equation (28) therein with $(1+ m \lambda_i) dt \to 0$, $\sigma^2=4 \gamma$ and $\beta=2$. The sum over $\beta\neq\al$ is clearly a manifestation of the typical Coulombic repulsion between eigenvalues. On top of that, the form of the noise is non-diagonal, as a combined effect of exactly preserving the trace and attributing Born's probability to trajectories. Then, in order to simplify the dynamics, we introduce a new set of unconstrained variables. 

\subsection{Mapping to Unconstrained Variables} \label{subsec:unconstrained_var}
As the dynamics of the eigenvalues exactly preserves the traces, it is useful to write them as $ \lambda_\alpha \equiv y_\alpha/(\sum_{\beta=1}^n y_\beta)$ in terms of a new set of variables $\{y_\alpha\}_{\alpha=1}^n$. Of course, the mapping between the $\lambda$'s and the $y$'s is not one-to-one, as a global rescaling of all the $y$'s does not affect the mapping. This freedom can be used to simplify the evolution of the $y$'s, in particular, by writing 
{
\begin{equation} \label{eq:dy}
    dy_\al = 4\gamma dt \, F_\al (\vec{y})
     + \sqrt{4\gamma} \, y_\al dB_\al \,,
\end{equation}
}
we can look for a force term $F_\alpha(\vec y)$ which is homogeneous of degree $1$. From standard It\^o calculus, one can verify that 
\begin{equation}
    F_\al(\vec{y}) = \frac x 4\sum_\beta y_\beta + \frac{y_\al^2}{\sum_\beta y_\beta} + \sum_{\beta\neq\al} \frac{y_\al y_\beta}{y_\al - y_\beta}+ g(\vec y) y_\alpha \,,
\end{equation}
where $g(\vec y)$ is an arbitrary homogeneous function ($g(c \vec y) = g(\vec y)$), which correctly reproduces Equation~\eqref{eq:eigen_x} for the evolution of the $\lambda$'s. Equation~\eqref{eq:dy} has the advantage of involving only a diagonal noise term, which is nonetheless multiplicative.
We can further simplify its form setting $w_\alpha = \log y_\alpha$, which leads to
\begin{equation} \label{eq:langevin_n}
    dw_\al = 4\gamma \tilde{F}_\al(\vec w)dt + \sqrt{4\gamma} dB_\alpha  \,,
\end{equation}
where $\tilde{F}_\al(\vec w) = e^{-w_\alpha} F_\alpha(e^{w})$. Setting $g(\vec y) = (n-1)/2$, one can express 
\begin{equation}
    \tilde F_{\alpha} = - \partial_{\alpha} V(\vec w) + \frac{x}{4}\mathsf{f}_\alpha (\vec w) \,,
\end{equation}
where $\partial_\alpha = \partial/\partial_{w_\alpha}$, the potential $V(\vec \omega)$ is
\begin{equation}
\label{eq:V}
    V(\vec \omega) = - \frac{1}{2} \sum_{\alpha \neq \beta} \log \sinh \frac{|w_\alpha - w_\beta|}{2} - \log \sum_\beta e^{w_\beta}  
\end{equation}
whereas the non-conservative force is given by $\mathsf{f}_\alpha = \sum_{\beta} e^{w_\beta-w_\al}$.
We will now explain how the dynamics induced by Equation~\eqref{eq:langevin_n} can be solved in different regimes.

\section{Stationary State at \boldmath{$x>0$} and \boldmath{$n\to\infty$}}
We now consider the dynamics induced by imperfect measurements at finite $x = O(1)$ in the large $n$ limit. When the dephasing term dominates the right-hand side of Equation~\eqref{eq:eigen_x}, the eigenvalue distribution is expected to be peaked around $\lambda = 1/n$. Therefore, it is convenient to rescale the eigenvalues via 
\be \label{eq:lambda_tilde}
\tilde \lambda_\al = \frac{2n\lambda_\al}{x} \,,
\ee
with the trace condition becoming $\sum_\al \tilde \lambda_\al= 2n/x$.  
With this scaling, we can analyze the noise term in Equation~\eqref{eq:eigen_x}, which is proportional to $\tilde\lambda_\alpha( dB_\alpha - x/(2n) \sum_\beta \tilde\lambda_\beta\, dB_\beta ) \sim \tilde \lambda_\alpha \,dB_\alpha$. Indeed, by summing the last term in quadrature, we see it scales like $n^{-1/2}$ and can thus be neglected for $n\gg 1$. This can be understood
as the sum $\sum_\al\tilde \lambda_\al$ being extensive and in the $n \rightarrow \infty$ limit its fluctuations are subleading with respect to its average. In this scaling, Equation~\eqref{eq:eigen_x} becomes 
\begin{equation}
\label{eq:eigen_x_tilde}
    d\tilde \lambda_\al = 4\gamma dt \left[ \sum_{\beta\neq\al} \frac{\tilde \lambda_\al\tilde \lambda_\beta}{\tilde \lambda_\al - \tilde \lambda_\beta} + \frac n 2 \left( 1-\frac x 2 \tilde \lambda_\al \right) \right]  + \sqrt{4\gamma} \, \tilde \lambda_\al  dB_\al\,.
\end{equation}
This associated stochastic dynamics is exactly solvable even at finite $n$. 
Indeed, by introducing again logarithmic variables $\tw_\al = \log \tilde \lambda_\al$, we can reduce this to a Langevin equation, analogous to Equation~\eqref{eq:langevin_n}, as follows:
\begin{equation} \label{eq:langevin_nw}
    d\tw_\al = 4 \gamma dt (-\partial_\al V_x( \vec\tw)) + \sqrt{4\gamma}\, dB_\al \,.
\end{equation}
We stress that this equation involves a completely conservative force where the potential $V_x(\vec \tw)$ now reads as follows: 
\begin{equation}
    V_x (\vec \tw) = - \frac 1 2 \sum_{\al \neq \beta} \log \sinh \frac{|\tw_\al - \tw_\beta|}{2} + \frac n 2 \sum_\al \left[ e^{-\tw_\al} + \left( 1+\frac x 2 \right) \tw_\al \right] \,.
\end{equation}
It is helpful to comment between the relation of the variables $\tw_\alpha$ introduced here and $w_\alpha$ in Section\,\ref{subsec:unconstrained_var}. 
We can see that they match, up to an inessential shift of the center of mass. The initial value of the sum $\sum_\al e^{w_\al}$ can be chosen arbitrarily and in the limit $n \rightarrow \infty$ it becomes a conserved quantity. Since we implicitly chose $\sum_\al e^{\tw_\al} = \sum_\al \tilde\lambda_\al = 2n/x$, we can rewrite  $\lambda_\al = e^{\tw_\al}/\sum_\al e^{\tw_\al}$ coherently with Section\,\ref{subsec:unconstrained_var}. 

Thanks to the Langevin form of Equation~\eqref{eq:langevin_nw}, we can easily deduce the joint distribution
\begin{equation} \label{eq:dens_mat_inv_wish_w}
    P_{\mathrm{stat}}(\vec \tw) = \frac{1}{Z} e^{-2V(\vec \tw)} = \frac{1}{Z} \prod_{\alpha > \beta} \sinh \Bigl(\frac{\tw_\alpha - \tw_\beta}{2}\Bigr)^2 \prod_{\alpha} e^{-\tw_\alpha n(1+ \frac{x}{2})} e^{-n e^{-\tw_\alpha}} \,, 
\end{equation}
which is the stationary solution to the associated FP equation. In terms of the rescaled eigenvalues $\tilde\lambda_\al$ 
\begin{equation} \label{eq:dens_mat_inv_wish}
    P_{\mathrm{stat}}(\vec{\tilde\lambda}) 
    =  \frac{1}{Z} \frac{\prod_{\alpha > \beta} (\tilde \lambda_{\alpha} - \tilde \lambda_{\beta})^2}{ \ \prod_\alpha \tilde\lambda_\alpha^{n(2+x/2)}}  e^{-n \sum_\al \tilde \lambda^{-1}_\alpha} \,,
\end{equation}
which is, upon rescaling $\tilde \lambda_\al \to \tilde \lambda_\al/2n = \lambda_\al/x$, the joint distribution function for the eigenvalues of a matrix in the inverse-Wishart ensemble, with $m=n(1+x/2)$. We refer to Appendix~\ref{inverseWishart} for a brief discussion of such an ensemble. 

\begin{figure}
    \includegraphics[width=\textwidth]{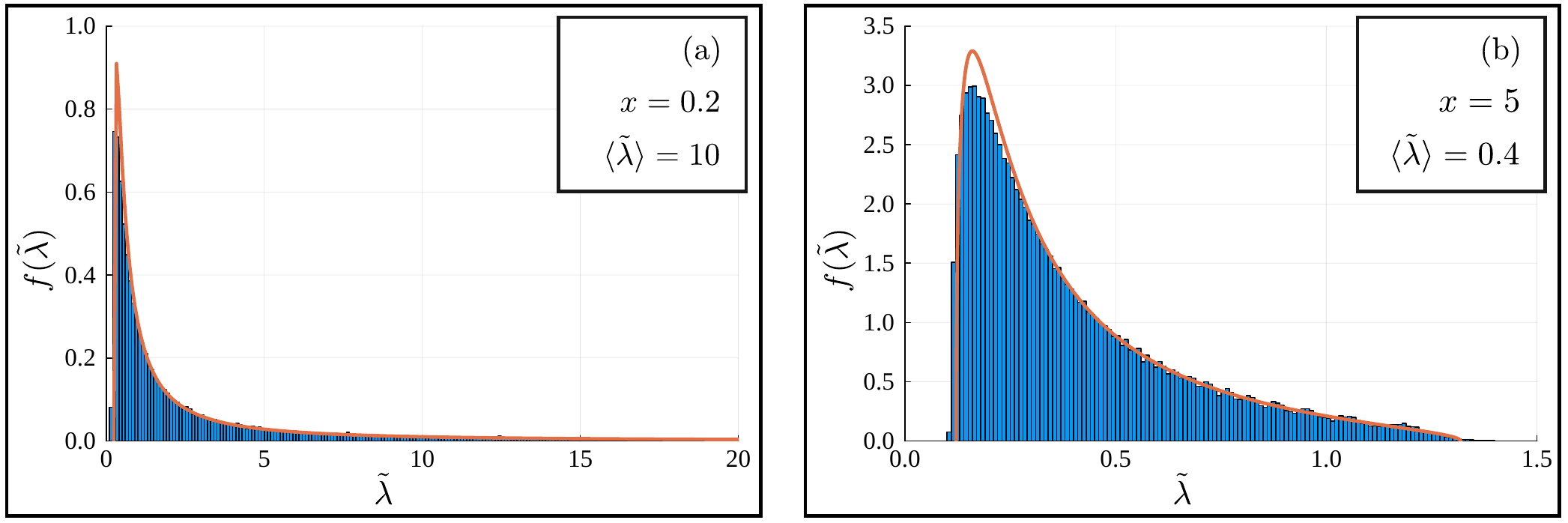}
    \caption{Inverse-Marchenko--Pastur
 distribution. The spectral density $f(\tilde \lambda)$ for the rescaled eigenvalues $\tilde \lambda$ defined in Equation~\eqref{eq:lambda_tilde} is shown. In the large $n$ limit, it takes the inverse-Marchenko--Pastur form given in Equation~\eqref{eq:f_IMP_rho}. In both plots, the orange line displays the theoretical curve, whereas the blue histogram bars are computed after a numerical simulation (with $n=50$) of the weak measurement protocol. (\textbf{a}) The spectral density at $x=0.2$ in the range $[0,4/x]$. At small $x$, most eigenvalues are located in vicinity of $\tilde \lambda_-\to 1/4$, as an effect of purification. (\textbf{b}) The spectral density at $x=5.0$ in its domain $[\tilde\lambda_-,\tilde\lambda_+]$. For larger values of $x$, the rescaled eigenvalues take finite values around their average $\braket{\tilde\lambda}=2/x$.}
    \label{fig:IMP}
\end{figure}

Let us now introduce the one-particle density function 
\begin{equation}
    f(\tilde \lambda) = \frac{1}{n} \sum_\al \delta(\tilde \lambda-\tilde \lambda_\al)\,,
\end{equation}
satisfying $\int d\tilde \lambda f(\tilde \lambda)=1$, $\int d\tilde \lambda f(\tilde \lambda) \tilde \lambda=2/x$. 
For matrices in the inverse-Wishart ensemble, the eigenvalue density takes the inverse-Marchenko--Pastur (IMP) expression reported in Equation~\eqref{eq:f_IMP} of Appendix~\ref{inverseWishart}. 
It is thus immediate to verify 
that the distribution for the rescaled eigenvalues defined above reads
\begin{equation} \label{eq:f_IMP_rho}
    f(\tilde\lambda, x) \equiv f_{\mathrm{IMP}}(\tilde\lambda,x) =\frac{x}{4\pi \tilde\lambda^2} \sqrt{(\tilde\lambda-\tilde\lambda_-)(\tilde\lambda_+-\tilde\lambda)} \,,  
\end{equation}
and it is shown in Figure~\ref{fig:IMP}. 
The endpoints $\tilde\lambda_\pm$ are given by
\begin{equation} \label{eq:f_IMP_endpoints}
    \tilde\lambda_{\pm} = \frac{4}{x^2}\left( \sqrt{1+\frac x 2}\pm1 \right)^2 \,.
\end{equation}
The distribution in Equation~\eqref{eq:f_IMP_rho} is valid in the $\hilbn\to\infty$ limit
which is also needed for Equation~\eqref{eq:eigen_x_tilde}
to become a valid approximation of the original~Equation~\eqref{eq:eigen_x}.
The behavior of $f_\mathrm{IMP}(\tilde\lambda,x)$ as $x$ is varied and defines a sharply peaked distribution around $2/x$ for $x\gg 1$, as the interval $\tilde\lambda_+-\tilde\lambda_-=\frac{16}{x^2}\sqrt{1+\frac x 2}$ collapses to a point. Conversely, in the limit of nearly perfect measurements $x \ll 1$, the rescaled eigenvalues spread on the positive semi-axis between $\tilde\lambda_-\to1/4$ and $\tilde\lambda_+\to\infty$, whereas $f(\tilde\lambda)$ remains peaked around $\tilde\lambda = 1/3$. As this corresponds to $\lambda \rightarrow 0$, this signals the purification dynamics induced by perfect measurements. For any finite $n$, it is important to notice, however, that even in this regime, one has to assume $x \gg n^{-1}$ for our approximation to be valid, so that the limits $x \rightarrow 0$, $n \rightarrow \infty$ do not commute.  

The Von Neumann entropy, defined in Equation~\eqref{eq:VN_entropy_def}, can be computed in the stationary state, in terms of the inverse-Marchenko--Pastur eigenvalue distribution above, as follows:
\begin{equation} \label{eq:VN_entropy_x}
    \braket{S_1}_{\mathrm{stat}} = S_{\mathrm{max}} - \log\frac x 2 - \frac x 2 \int d\tilde \lambda \, f_{\mathrm{IMP}}(\tilde \lambda) \,\tilde \lambda \log \tilde \lambda \,, 
\end{equation}
with $S_{\mathrm{max}} = \log n$ the entropy of the maximally mixed state. The last integral on the right-hand side can be evaluated by means of contour integration
\begin{equation}
    - \frac x 2  \frac{x}{4\pi} \int_{\tilde\lambda_-}^{\tilde\lambda_+} d\tilde\lambda \frac{\log \tilde\lambda}{\tilde\lambda} \sqrt{(\tilde\lambda-\tilde\lambda_-)(\tilde\lambda_+-\tilde\lambda)} =  1 + \left( 2+\frac x 2 \right) \log \frac x 2 - \left( 1 + \frac x 2 \right) \log\left( 1 + \frac x 2\right) \,, 
\end{equation}
so that one has the remarkably simple exact formula, as follows: 
\begin{equation}
   \braket{S_1}_{\mathrm{stat}} - S_{\mathrm{max}} =  1 - \left( 1 + \frac x 2 \right) \log\left( 1 + \frac 2 x  \right) \,.
\end{equation}
In particular, this allows us to evaluate its asymptotic behavior in the opposite regimes $x\gg1$ and $n^{-1} \ll x \ll 1$, as follows:
\begin{equation}
    \braket{S_1}_{\mathrm{stat}} - S_{\mathrm{max}} = 
    \begin{cases}
        -\frac 1 x + O(x^{-2}) \,,& x \gg 1 \\
        -\log(x)+O(1)  \,,& x \ll 1\
    \end{cases} \,.
\end{equation}

\subsection{Finite Time Dynamics} \label{sec:finitexfinitet}

The finite time evolution of the set of unconstrained $\{\tw_\al\}_{\al=1}^n$ used in this section has been studied in detail by one of the authors in Ref.~\cite{gautie2021matrix}. Our Langevin equation \eqref{eq:langevin_nw} indeed reduces to Equation~(42) of Ref.~\cite{gautie2021matrix} upon the identification $t\to\Gamma x t$, $m=-1$, $\sigma=\sqrt{4/xn}$, $\beta=2$. Consistent with our result in Equation~\eqref{eq:f_IMP_rho}, a stationary inverse-Wishart distribution was also obtained there.
Here, we briefly summarize the procedure and the results with our notation and parameters. The associated FP equation can be mapped to an imaginary time quantum problem via the transformation 
\begin{equation}
\label{eq:FPtoSchr}
   P(\vec \tw,t) = [P_{\mathrm{stat}}(\vec \tw)]^{1/2} \,\Psi(\vec \tw,t) \,,  
\end{equation}
with $P_{\mathrm{stat}}(\vec \tw)$ given in Equation~\eqref{eq:dens_mat_inv_wish_w} (we better discuss the quantum mapping from FP to Schr{\"o}dinger in Section~\ref{subsec:exact_finite_time}). The ensuing Schr{\"o}dinger equation reads
\begin{equation}
    \partial_t \Psi = - 2\gamma [\mathcal{H}-E_0] \Psi \,, 
\end{equation}
with the Hermitian Hamiltonian
\begin{equation}
    \mathcal H = \sum_\al \left[ -\frac{\partial^2}{\partial \tw_\al^2} + U(\tw_\al) \right]\,, 
\end{equation}
and the finite energy shift $E_0$, which is the $n$-particle ground state of $H$. As the trajectories of the $n$ particles are non-crossing, the problem is mapped onto fermions. The one-particle potential $U(\tw_\al)$ is of Morse type:
\begin{equation} \label{eq:morse_potential}
    U(\tw_\al) = \frac{(xn)^2}{4}\left[ \frac{e^{-2\tw_\al}}{4}- \left( \frac 2 x + \frac 1 2 \right) e^{-\tw_\al} \right] \,. 
\end{equation}
The single-particle spectrum of the Morse potential features a finite number of bound states $\psi_k(\tw)$ at eigenenergy $E_k=-[k+1/2-n(1+x/4)]^2$, with $k\in\mathbb{N}$, $k<k_\mathrm{max}$, $k_\mathrm{max}<\lfloor n(1+x/4)-1/2 \rfloor$, and a continuum of scattering states $\phi_p(\tw)$, such that $E_p=p^2$, with $p\in\mathbb{R}$. 
Let us notice that the number of bound states is always greater than the number of particles, i.e., $\lfloor n(1+x/4)+1/2 \rfloor>n\,\forall x>0$, and the scattering states are empty. The $n$-fermion ground state is thus $E_0 = \sum_k E_k$. 
Consequently, the $n$-fermion energy states of the original FP operator are simply given by 
\begin{equation}
    \varepsilon(n) = \frac{2}{xn}[E(n)-E_0]\,,
\end{equation}
so that the lowest eigenvalue is $\varepsilon_0(n)=0$. 
The single-particle finite time evolution from an initial $w_0$ to a final $w$ coordinate is determined by the Euclidean propagator $G_{\rm M}$ using the spectral decomposition over the bound and the scattering states
\begin{equation}
    G_{\rm M}\left(\tw,\tw^0, \tau\right) := \sum_{k=0}^{k_\mathrm{max}} \psi_k(\tw)\psi_k(\tw^0)e^{-E_k \tau} +\int_\mathbb{R} \frac{dp}{2\pi} \phi_p(\tw)\phi_p(\tw^0)e^{-p^2 \tau}\,.
\end{equation}
Then, the $n$-fermion propagator is obtained in terms of the Karlin--McGregor determinant formula \cite{pjm/1103038889} for the probability evolution of $n$ non-intersecting particles from $\vec \tw^0$ to $\vec \tw$. From the quantum Euclidean propagator, one can write the solution to the original FP problem using~Equation~\eqref{eq:FPtoSchr} as
\begin{equation}
\label{eq:Pfinitetx}
    P \left(\vec \tw, t;\vec \tw^0,t=0 \right) = e^{E_0\Gamma x t} \,\frac{P^{1/2}_{\mathrm{stat}}(\vec \tw)}{P^{1/2}_{\mathrm{stat}}(\vec \tw^0)}   
    \det_{1\leq \al,\beta \leq n}  
    G_{\rm M}(\tw_\al,\tw^0_\beta,\Gamma x t) \,,
\end{equation}
given the initial condition $P(\vec \tw,t=0) = \prod_\al \delta(\tw_\al-\tw_\al^0)$.

\section{The Perfect Measurement Dynamics}

In this section, we present the full solution of the joint eigenvalue distribution for the measurement problem at arbitrary $n$ and time $t$, setting $x=0$.

\subsection{Exact Solution at Finite Time} \label{subsec:exact_finite_time}

The stochastic Equation \eqref{eq:langevin_n} takes the Langevin form
\be 
dw_\alpha = 4 \gamma dt ( - \partial_\alpha V ) + \sqrt{4 \gamma} dB_\alpha  \,,
\ee 
with the potential in Equation~\eqref{eq:V}. The joint distribution $P(\vec w,t)$ for the variables $w$'s satisfies the FP equation 
\be 
\partial_t P(\vec w,t) = 2 \gamma \sum_\alpha \partial_\alpha \left[\left( \partial_\alpha + 2 \partial_\alpha V \right) P(\vec w,t) \right]  \,.
\ee 
Formally, one can obtain a stationary solution to the FP equation, taking {$P_{\mathrm{stat}}=e^{-2 V}$}.
Of course, this is not normalizable, consistent with the fact that at $t \to \infty$ the density matrix simply purifies. Nonetheless, it can once again be used to convert the FP into a quantum problem, similarly to what was conducted in Section~\ref{sec:finitexfinitet} with Equation~\eqref{eq:FPtoSchr}. One has
\begin{equation}
\label{eq:schrFP}
P = \Psi {P_{\mathrm{stat}}^{1/2}} \quad \Rightarrow \quad  \partial_t \Psi =  - 2 \gamma\mathcal{H} \Psi \quad \,,
\end{equation} 
which does satisfy a Schr\"odinger evolution with Hamiltonian operator
\begin{equation}
\mathcal{H} =  \sum_\alpha \big[ - \partial_\alpha^2 
+  (\partial_\alpha V)^2 - \partial^2_\alpha V \big] \,.
\end{equation} 
Remarkably, after using some algebra (see Appendix \ref{identities} and also \cite{10.21468/SciPostPhys.11.6.110, Ipsen_2016}), one can show that 
\begin{equation}
   E_n \equiv (\partial_\alpha V)^2 - \partial^2_\alpha V = \frac{1}{12} \left(n^3+11 n\right) \,,
\end{equation}
implying that the Schr\"odinger evolution in Equation~\eqref{eq:schrFP} amounts to free diffusion with the additional non-crossing constraint when two $w$'s collide, which maps on free fermions, as the external Morse potential Equation~\eqref{eq:morse_potential} present in \cite{gautie2021matrix}, see Equation (47), is absent here. The propagator for such dynamics can be obtained using the Karlin--McGregor determinant formula~\cite{pjm/1103038889}, also used in Equation~\eqref{eq:Pfinitetx}. For a given initial condition $P(\vec w, t = 0) = \prod_\alpha \delta(w_\alpha - w_\alpha^0)$, we 
can obtain the solution 
\begin{equation}
    P\left(\vec w, t; \vec w^0, t=0\right) = { \frac{P_{\mathrm{stat}}^{1/2}(\vec w)}{P_{\mathrm{stat}}^{1/2}(\vec w_0)} } (e^{- 2 \gamma H t})_{\vec w , \vec w^0} \,,
\end{equation} 
which leads to 
\begin{equation}
\label{eq:fullprop}
    P\left(\vec w, t; \vec w^0, t = 0\right) = e^{-2 \gamma E_n t} \, \frac{\sum_\alpha e^{w_\alpha}}{ \sum_\alpha e^{w^0_\alpha}} \prod_{\alpha < \beta} \frac{ \sinh  \frac{|w_\alpha - w_\beta|}{2}}{\sinh \frac{|w^0_\alpha - w^0_\beta|}{2}} \det_{1 \leq \alpha, \beta \leq n} G_0(w_\alpha, w^0_\beta,2 \gamma t) \;.
\end{equation}
In this expression, we introduced the free diffusion propagator, namely $G_0(w,w',\tau) = \exp[ - \frac{(w-w')^2}{4 \tau}] / \sqrt{4 \pi \tau}$. In the following, we focus on the case of the infinite-temperature initial state, where all the initial conditions coincide $\vec w^0 \to 0$ (the specific value is inessential). In that limit, Equation~\eqref{eq:fullprop} further simplifies as
$P(\vec w, t) \to  \frac{1}{Z_t}P_0(\vec w, t) \left( \sum_\alpha e^{w_\alpha} \right)$, where 
\begin{equation} 
\label{eq:P0expr}
    P_0(\vec w, t) = \frac{1}{Z^0_t} \prod_{\alpha < \beta} \sinh \frac{|w_\alpha - w_\beta|}{2} \prod_{\alpha < \beta} |w_\alpha-w_\beta| \,e^{- \sum_\alpha \frac{w_\alpha^2}{8 \gamma t}} \,,
\end{equation}
and the time dependent constants $Z^0_t, Z_t$ are both determined by enforcing the normalization of $P_0(\vec w, t)$ and $P(\vec w, t)$, respectively. 

\subsection{Relation between the Two Averages}

The distribution $P_0(\vec w, t)$ has appeared in~\cite{Ipsen_2016, Mergny_2021} in studying the dynamics of the so-called isotropic Brownian motion in the context of May-Wigner instability. In contrast, the expression for $P(\vec w, t)$ has the extra factor  $\sum_\alpha e^{w_\alpha}$, whose origin can be traced back to Born's rule, expressing (up to an irrelevant normalization) the probability in Equation~\eqref{eq:rhoa} as $p_\aaa \propto \sum_\alpha e^{w_\alpha}$. It is useful to  identify the 
\begin{equation}
\label{eq:spectrum}
    \mbox{spectrum}[\tilde\rho] = \{e^{w_\alpha}\}_{\alpha = 1}^n \,,
\end{equation}
 where $\tilde\rho$ is the unnormalized density matrix introduced in Equation~\eqref{eq:rhoa}. In analogy with other problems of a multiplicative process involving random matrices~\cite{derrida_singular_1983,bouchard_rigorous_1986,ipsen_products_2015,haake1991quantum}, we will refer to the variables $w_\alpha$ as the Lyapunov exponents. In the following, consistent with the definition in Equation~\eqref{eq:forcemeas}, we will add a subscript $0$ to the averages computed with $P_0$, with the general relation
\begin{equation}
\label{eq:F0ave}
\langle F(\vec w) \rangle = 
\frac{
\langle F(\vec w) \left( \sum_\alpha e^{w_\alpha} \right) \rangle_0}{\langle \sum_\alpha e^{w_\alpha} \rangle_0}
\end{equation}
for an arbitrary function $F(\vec w)$. 
For instance, for the Renyi's entropies we can write the following: 
\begin{align}
    &\langle S_k \rangle = 
    \frac {\langle \tr \tilde \rho \rangle_0^{-1}} {1-k} \left\langle \log\left[\frac{\tr \tilde\rho^k}{(\tr\tilde \rho)^k}\right] \tr \tilde \rho \right\rangle_0   \;,  \label{eq:renyitrho} \\
    &\langle S_1 \rangle = 
    - \langle \tr \tilde \rho \rangle_0^{-1}\left\langle \left[\frac{\tilde\rho}{\tr \tilde \rho} \log\frac{\tilde\rho}{\tr \tilde \rho}\right]  \tr \tilde \rho \right\rangle_0     \;.
    \label{eq:VNtrho}
\end{align}    
Then, using~Equation~\eqref{eq:spectrum}, we can rewrite the Von Neumann entropy as 
\begin{equation}
\label{eq:S1full}
    \langle S_{1} \rangle 
      = \frac{\left\langle \sum_\alpha e^{w_\alpha} \ln\sum_\beta e^{w_\beta} \right\rangle_0}{\langle \sum_\alpha e^{w_\alpha} \rangle_0} - \frac{
      \left\langle \sum_\alpha e^{w_\alpha} w_\alpha \right\rangle_0}{\langle \sum_\alpha e^{w_\alpha} \rangle_0} \,.
\end{equation}
In particular, introducing the moments of the eigenvalues and of the trace as
\label{eq:momdef}
\begin{align}
\label{eq:momMdef}
    &M(m) := \left\langle \sum_\alpha e^{m w_\alpha} \right\rangle_0 \,, \\
    &\Omega(m) := \left\langle \Big(\sum_\alpha e^{w_\alpha} \Big)^m \right\rangle_0 \,,
    \label{eq:momtrdef}
\end{align}
we can express Equation~\eqref{eq:S1full} as 
\begin{equation} \label{eq:VNrepl}
 \langle S_1 \rangle = \frac{1}{M(1)} \partial_m \left.(\Omega(m) - M(m))\right|_{m=1} = \left.\partial_m \log \frac{\Omega(m)}{M(m)} \right|_{m=1} \,.
\end{equation}
Thus, below we will study these moments with the measure $P_0$. We observe that in Equation~\eqref{eq:P0expr}, one can recognize two Vandermonde determinants, since
\label{eq:vdmgen}
\begin{align} 
&\Delta(\vec w) \equiv \prod_{\alpha < \beta} (w_\alpha - w_\beta) = \det(w_\alpha^{k-1})_{k,\alpha=1}^n \,, \label{eq:vdm1} \\
& \prod_{\alpha < \beta} [2\sinh(\frac{w_\alpha - w_\beta}{2})] = \det(e^{\delta_k w_\alpha})_{k,\alpha=1}^n \,, \label{eq:vdm2}
\end{align}
where we set $\delta_k = (n + 1)/2 - k$. This implies that $P_0$ describes a determinantal point process, which allows us to obtain several exact results, as discussed below. 

\section{Exact Results for the Unbiased Ensemble}

\subsection{Average of Schur's Polynomials}

The calculation of several quantities, including Renyi's entropies \eqref{eq:renyidef}, requires the expressions of correlation functions of the Lyapunov exponents $w$'s. Here, we explain how they can be expressed systematically. First, because of the symmetry under exchange in the $w$'s, we can restrict to symmetric functions. Then, because of the determinantal structure in Equation~\eqref{eq:vdmgen}, following~\cite{forrester2020global}
a complete and useful basis of symmetric functions is given by Schur's polynomials and each correlation function can be expressed once the expectation of Schur's polynomials is known.
To a partition $\kappa = (\kappa_1,\ldots, \kappa_n)$ of the integer $m = \sum_j \kappa_j$, with
$\kappa_1\geq\kappa_2\geq\ldots\geq\kappa_n \geq 0$, one associates
the corresponding Schur polynomial via~\cite{macdonald1998symmetric}
\begin{equation}
    s_\kappa(y) = \frac{\det(y_\alpha^{\kappa_j + n - j})_{j,\alpha=1}^n}{\det(y_\alpha^{k-1})_{k,\alpha=1}^n} \;.
\end{equation}
Setting $y_\alpha = e^{w_\alpha}$ and denoting $h_j = \kappa_j + n - j$, we can now express the average
\begin{equation}
\label{eq:schurdet}
    \langle s_\kappa(y) \rangle_0 = C A_t^m \int d\vec w \, \Delta(\vec w)
    \det(e^{h_j w_\alpha}) \, e^{-\sum_\alpha \frac{w_\alpha^2}{8 \gamma t}} \,, 
\end{equation}
where $C$ is the normalization and {the} constant $A_t$ raised to the power $m$ accounts for the shift of the center of mass and will be fixed below.
We can use Andreief identity~\cite{doi:10.1142/S2010326319300018} to express it in terms of a single determinant
\begin{equation}
    \langle s_\kappa(y) \rangle_0 = C A_t^m
    \det( I_{k, h_j})_{k,j=1}^n \;,
\end{equation}
where we defined
\begin{equation}
    I_{k,h} = \int_{-\infty}^\infty \frac{dw}{\sqrt{8 \pi t \gamma}} w^{k-1}  e^{ h w - \frac{w^2}{8 \gamma t}} = \left.\partial_{\mu}^{k-1} \left[ e^{2 t \gamma \mu^2} \right] \right|_{\mu = h} \,.
\end{equation}
We can thus express the coefficients $I_{k,h}$ in terms of the Hermite polynomials
$H_n(x) = (-1)^n e^{x^2} \partial_x^n [e^{-x^2}]$ as
\begin{equation}
    I_{k, h} =e^{2 h^2 \gamma t} H_{k-1}\left(i \sqrt{2 t \gamma } h\right) \,.
\end{equation}
where again we absorbed some extra constants by redefining $C$. The fact that at large $x$, $H_\ell(x) = 2^\ell x^\ell + O(x^{\ell -1 })$ can be used to express the determinant
\begin{equation}
\det[ I_{k, \kappa_j +j}]_{k,j=1}^n
\propto \exp\left[2 \gamma t \sum_j h_j^2\right] \det[h_j^{k-1}] \,.
\end{equation}
This last determinant is once again a Vandermonde one which can be expressed via \eqref{eq:vdm1}. We can now plug this back into Equation~\eqref{eq:schurdet} 
and fix the constant $C$ using $s_{\kappa = 0}(y) = 1$, where $h_j \to n - j$. We finally obtain
\begin{equation}
\label{eq:schurave}
\langle s_\kappa(y) \rangle_0 = A_t^m
e^{2 \gamma t \sum_{j=1}^n (h_j^2 - (j-1)^2)}s_\kappa(1) \,,
\end{equation}
where we recognized the equality
\begin{equation}
\label{eq:schur1}
\prod_{1 \leq j<j' \leq n}\frac{h_j - h_{j'}}{j'-j}  = s_\kappa(y_1 = 1, \ldots, y_\hilbn = 1) \,,
\end{equation}
which expresses the number of semistandard Young diagram of shape $\kappa$ and $n$ entries~\cite{macdonald1998symmetric}.
We can now use $\sum_{j=1}^n h_j^2 - (j-1)^2=  (2n - 1) m +2 \nu(\kappa)$~\cite{macdonald1998symmetric}, with
\begin{equation}
\label{eq:nudef}
\nu(\kappa) = \sum_{j} \binom{\kappa_j}{2} - \binom{\kappa_j'}{2} \,,
\end{equation}
where $\kappa' = (\kappa_1', \kappa_2',\ldots)$ denotes the partition dual to $\kappa$, with $\kappa'_i = \# \{ \kappa_j | \kappa_j \geq i\}$. The above average can then be expressed as follows:
\begin{equation}
\label{eq:schuravesimple}
\langle s_\kappa(y) \rangle_0 = A_t^m e^{2\gamma t m (2n-1)} e^{4 \gamma t \nu(\kappa)} s_\kappa(y_\alpha = 1) \,.
\end{equation}

\subsection{Power-Law Symmetric Polynomials}

For later convenience, we also introduce the power-law symmetric polynomials
\begin{equation}
    p_j(y) = \sum_\alpha y_\alpha^j \,,
\end{equation}
and additionally, for any integer partition $\mu = (\mu_1,\ldots,\mu_n)$, we can set
\begin{equation}
    p_{\mu}(y) = \prod_j p_{\mu_j}(y) \,,
\end{equation}
which form a complete linear basis for symmetric polynomials. For instance, we can express
\begin{equation}
    p_{1^m}(y) = \left(\sum_\alpha y_\alpha\right)^m \,.
\end{equation}
Since Schur's polynomials are also a complete basis, one can find a change in basis between the two. It is given as (see Equation (3.10) in \cite{10.1063/5.0048364,forrester2020global})
\begin{equation} \label{eq:symmetric_to_schur}
    p_{\mu}(y) = \sum_{\kappa\vdash m} \chi^{\kappa}_\mu s_{\kappa}(y) \,,
\end{equation}
where $\chi_\mu^\kappa$ represents a character of the symmetric group $S_m$, with $m=\sum_\alpha \kappa_\alpha$, on the irreducible representation and the conjugacy class labeled, respectively, by the integer partitions $\kappa$ and $\mu$. 

\subsection{Calculation of the Moments}

Thanks to the relation between power-law symmetric polynomials and Schur's ones, given in Equation~\eqref{eq:symmetric_to_schur}, 
we can use the previous result to express the moments of eigenvalues, c.f. Equation~\eqref{eq:momMdef} with $y_\alpha = e^{w_\alpha}$:
\begin{equation}
\label{eq:momdefschur}
M(m) = \sum_{r=0}^{m-1} (-1)^r \langle s_{(m-r, 1^r)}(y)\rangle_0 \,,
\end{equation}
which corresponds to Equation~\eqref{eq:symmetric_to_schur} in the case $\mu = (m)$, i.e., a  single cycle of length $m$. In this case, the only non-vanishing characters $\chi_{(m)}^\kappa$ are those related to $L$-shaped partitions $\kappa = (m-r, 1^r) = (\kappa_1 = m-r, \kappa_2 = 1,\ldots, \kappa_{r+1}=1)$, with $r=0,\ldots,m-1$. In particular, one has $\chi_{(m)}^{(m-r,1^r)}=(-1)^r$. Therefore, from Equation~\eqref{eq:nudef}, we obtain $2\nu(\kappa)= m(m-2 r-1)$, and 
\begin{equation}
\label{eq:schurL}
    s_{(m-r, 1^r)}(1) = 
    \frac{\Gamma (m+n-r)}{m \Gamma (r+1) \Gamma (m-r) \Gamma (n-r)} \;.
\end{equation}
The sum in Equation~\eqref{eq:momdefschur} thus becomes
\begin{equation}
\label{eq:momfinN}
    M(m) = \,e^{2 \gamma t m(m+n)} \sum_{r=0}^{m-1} (-1)^r s_{(m-r, 1^r)}(1) \,e^{- 2\gamma t m (2r+1)} \,,
\end{equation}
which can be expressed for integer $n$ in terms of the hypergeometric function $\, _2F_1(a,b; c;z)$ \cite{enwiki:1177949416}, obtaining the following:
\begin{equation}
    \label{eq:momfinNsum}
    M(m) = \binom{m+n-1}{n-1} \,e^{2\gamma t m (m+n-1)}\, _2F_1(1-m,\,1-n;\,1-m-n;\,z)\,.
\end{equation}
In both expressions, we have fixed the value of $A_t = e^{-2\gamma t (n-1)}$ using $\partial_m M(0) = \langle \sum_\alpha w_\alpha \rangle_0 = 0$ in our conventions. 
Note that $M(0)=n$ as required, and that the lowest moments have simple expressions, e.g., one has $M(1)=\langle \sum_\alpha e^{w_\alpha} \rangle_0=n e^{2 n \gamma t}$.

Moreover, it is also possible to compute the moments of the trace defined in \mbox{Equation~\eqref{eq:momtrdef},} with $y_\alpha = e^{w_\alpha}$, by making use of the same relation.
We can express
\begin{equation} \label{eq:OmSchur}
    \Omega(m) = \sum_{\kappa\vdash m} \chi_{1^m}^\kappa \langle s_\kappa(y) \rangle_0 = e^{2t\gamma n} \sum_{\kappa \vdash m}
     e^{4 \gamma t \nu(\kappa)} \chi_{1^m}^\kappa s_\kappa(1) \,,
\end{equation}
where the sum is over the partition $\kappa$ of the integer $m$.

\subsection{Equivalent Formulations}

Equation~\eqref{eq:P0expr} can appear in different contexts that provide interesting interpretations.  Following~\cite{BREZIN1996697,PhysRevE.58.7176,johansson2001universality}, and using \eqref{eq:vdm1} and \eqref{eq:vdm2}, at a fixed time $t$, one can identify the Lyapunov exponents $w$'s with the spectrum of 
the matrix 
\be \label{eq:W} 
W = \sqrt{4 \gamma t n} H + (4 \gamma t) D \,,
\ee
where $H$ is drawn from the GUE distribution \eqref{eq:GUE} and $D = \operatorname{diag}(\frac{n-1}{2},\frac{n-3}{2},\ldots, -\frac{n-1}{2}) = \operatorname{diag}(\delta_1,\ldots, \delta_n)$, see also \cite{claeys2014random}. 
The scaled distribution of the largest eigenvalue of $W$ was computed at large $n$ in \cite{krajenbrink2021tilted}, see Section III there, and found to be GUE Tracy-Widom, since no localization transition takes place when the elements of $D$ are equispaced; see also \cite{claeys2018propagation}. Besides, equation~\eqref{eq:W} is particularly effective for numerical sampling from the distribution Equation~\eqref{eq:P0expr}.

Equivalently, the $w$'s can be seen as performing a DBM with $\beta=2$ in inverse time $\sim 1/t$. For time inversion of the DBM we refer to, e.g., Appendix B in \cite{krajenbrink2021tilted} with equally spaced initial condition.
Indeed, one can rewrite \eqref{eq:W} as
\be \label{X} 
W = 4 \gamma t n X  \, , \quad X = {\rm diag}\left(\frac{\delta_i}{n} \right)  + \sqrt{s} H   \,,
\ee 
with $s=1/(4\gamma t n)$, and the eigenvalues $\vec x(s)$ of $X$ have the same joint law
as a DBM at time $s$ with initial condition $\vec x(0)=\vec \delta/n$. This correspondence is given in Appendix~\ref{DBM}.
Note that the $\vec x(s)$ form a 
determinantal point process~\cite{johansson2001universality}, 
whose kernel is given in Appendix~\ref{app:kernel}. 
The interpretation of Equation~\eqref{eq:W} already indicates a qualitative behavior for the spectrum of the $w$'s: at initial times, the first term dominates and the distribution will resemble the one of a GUE in with eigenvalues in the support $[-4\sqrt{t \Gamma},4\sqrt{t \Gamma}]$. At larger times, the second term becomes more important and the distribution becomes uniformly spread in $[-2 t \Gamma, 2 t \Gamma]$.
It is also useful to write the distribution $P_0 \sim e^{-n^2\mathcal{E}_t(f)}$ as a functional of the single particle density 
\begin{equation}
\label{eq:densf}
    f(w) = \frac1n \sum_\alpha \delta(w - w_\alpha) \,,
\end{equation}
with the functional 
\begin{equation}
{\cal E}_t(f)  = \frac{1}{8 \Gamma t} \int dw \,w^2\, f(w)  -\frac{1}{2} \int dw dw' f(w) f(w') \left( \log \sinh \frac{|w-w'|}{2} + \log|w-w'| \right) \,.
\end{equation}
One can interpret this as the energy of gas particles in the presence of an external  harmonic trap that repel each other. When $\Gamma t = O(1)$, the different terms are of the same order: at large $n$, the particles will have a separation $\propto t \gamma / n$ and a continuous description emerges. We will analyze this regime in the next section. We will then discuss the long-time regime which emerges when $t \sim n/\Gamma = 1/\gamma$.

\subsection{Coulomb Gas Regime $\Gamma t\sim O(1)$}  

\begin{figure}
    \centering
    \includegraphics[width=\textwidth]{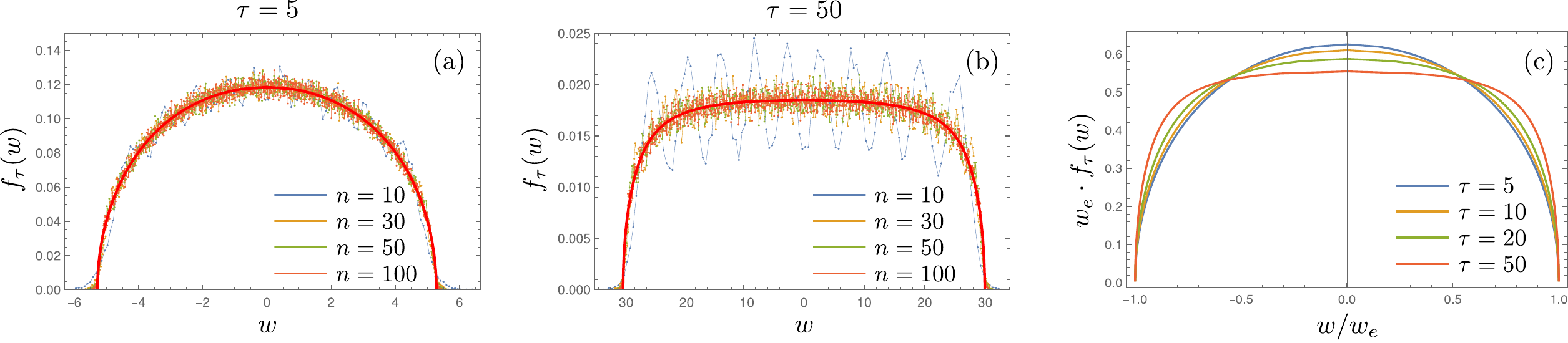}
    \caption{Short time behavior. The density $\dens_\tau(w)$ in the short time $t = \tau / 4\Gamma\sim 1/\Gamma$ regime, with $\tau$ finite. $\dens_\tau(w)$ features a crossover between a semi-circle at small $\tau$ and a square distribution at larger $\tau$. 
    (\textbf{a}) The small-$\tau$ semi-circle distribution is displayed for $\tau=5$ and increasing $n$ from $n=10$ to $n=100$. The red solid line shows the theoretical $n\to\infty$ curve.
    (\textbf{b}) The large-$\tau$ square distribution is shown for $\tau=50$ and increasing $n$, with the red solid line showing the $n\to\infty$ curve. 
    (\textbf{c}) The crossover from semi-circle towards square distribution is shown for increasing $\tau$ from $\tau = 5$ to $\tau=50$. All solid lines represent the theoretical density $w_ew \cdot\dens_\tau(w)$ on the rescaled $w/w_e$ axis, where $w_e$ is the edge coordinate for $\dens_\tau(w)$ in the $n\to \infty$ limit. }
    \label{fig:distribution_short_time}
\end{figure}

To quantitatively analyze the crossover between the semicircle and the uniform distribution predicted by the form Equation~\eqref{eq:W}, we rescale the time in the following way
\begin{equation} \label{scalingt} 
t = \frac{\tau}{4 \gamma \hilbn} = \frac{\tau}{4 \Gamma} \;,
\end{equation}
and consider the limit $n \to \infty$ at fixed $\tau$. 
The above correspondence in \eqref{X} shows that $s=1/\tau$, and $W=\tau X$.
The interpretation as a DBM in inverse time allows us to 
compute the limiting resolvent 
\be
g(z,\tau) := \lim_{\underset{t=\tau/(4 \gamma n) \to 0}{n \to +\infty}} \,  \frac{1}{n} \sum_i \frac{1}{z-w_i(t)} \,,
\ee 
using the complex Burgers equation (see details in Appendix \ref{DBM}). One finds that $g=g(z,\tau)$ satisfies the parametric equation
\be \label{zg} 
z =  \tau  g + \frac \tau 2 \coth(\frac{\tau g}{2}) \,.
\ee 
We can set $z=w-i 0^+$ and $g=g_r + i \pi f$, where $f$ is the spectral density. Eliminating the real part $g_r={\rm Re} (g)$ one finds (see Appendix \ref{DBM}) that the density $\dens=\dens_\tau(w)$ is the solution of
\begin{align}
\label{eq:densrho} 
& w = \pm \left[\log+ \left( b_\tau(\dens) + \sqrt{b_\tau(\dens) ^2-1} \right) + \frac{\pi \tau \dens}{\sin (\pi \tau \dens)} \sqrt{b_\tau(\dens)^2-1} \right] \,, \\
& b_\tau(\dens)  = \cos \pi \tau \dens + \frac{\sin \pi \tau \dens}{2  \pi \dens} \,.
\end{align}
Its support is an interval with edges at $w=\pm w_e = \pm \left[{\rm arccosh} (1+ \frac{\tau}{2}) + \sqrt{(1+\frac{\tau}{2})^2-1}\right]$.
As anticipated from the qualitative analysis of Equation~\eqref{eq:W}, the density interpolates (see Figure~\ref{fig:distribution_short_time}) between a semi-circle at small $\tau$, with $w_e \simeq 2 \sqrt{\tau} + \frac{\tau^{3/2}}{12}$, and a square distribution at large $\tau$ with $w_e = \frac{\tau}{2} + \log(e \tau) + \frac{1}{\tau}+  O(1/\tau^2) $. At all times $\tau>0$ the density vanishes at the edges as a square root $f(w) \sim B_{\tau} |w_e-w|^{1/2}$, with $B_{\tau}=\sqrt{2} / [\pi \tau^{3/4} (4 + \tau)^{1/4}]$, and Wigner--Dyson gap statistics near the edges. 
Although the density is known in parametric form, 
one can also use Equation~\eqref{eq:momfinN} to extract the exponential moments. After some manipulation on the hypergeometric function, we find
\begin{equation}
\mu(m) = \lim_{\substack{n \to \infty\\ t = \tau/n}} \frac{M(m)}{n} = \frac{e^{m \tau /2} L_{m-1}^{(1)}(-m \tau )}{m} \,,
\end{equation}
where $L_m^{(\alpha)}(x)$ denotes the generalized Laguerre polynomial, with $\lim_{m \to 0} \mu(m)=1$. A numerical check confirms that the density implicitly defined in Equation~\eqref{eq:densrho} satisfies
\begin{equation}
    \mu(m) = \int_{-w_e}^{w_e} dw f(w) e^{m w} \,.
\end{equation}
The fact that the variables become dense and described by a continuous distribution ensures that $\Omega(1)/n \to \mu(1)$ in Equation~\eqref{eq:momtrdef} is self-averaging and thus
\begin{equation}
    \lim_{n\to \infty} n^{-m} \Omega(m) = \mu(1)^m =  e^{\tau m/2}\,.
\end{equation}

\subsection{Universal Regime $\Gamma t = O(n)$ \label{sec:uni}}
\subsubsection{Scaling of the Edge}
Now, we consider the situation where time is scaled as $t = O(n/\Gamma)$. Using the rescaled rate $\gamma$, we can equivalently say that $\gamma t = O(1)$ whereas $n \to \infty$. In this regime
the repulsion due to the $\log \sinh$ term in  becomes more relevant. As a result, the Lyapunov exponents $w$ are well separated one from the other. More precisely, 
assuming the ordering $w_1 < w_2 <\ldots < w_n$,
one expects that the separation between two consecutive variables $w_{\alpha+1} - w_\alpha = O(\gamma t)$, so that the position of the edge $\bar{w}_{e} = \langle w_n \rangle_0 = O(t \gamma n)$. However, each variable $w_\alpha$ will have fluctuations of $O(\sqrt{\gamma t})$ so that the interactions are relevant and complicate the analysis. A similar lattice structure was already demonstrated in a related model involving a $\log \sinh$-potential~\cite{Forrester1994}. 
Here, the Coulombic repulsion $\log |w - w'|$ is an additional ingredient that modifies that short-distance behavior. We can obtain a preliminary understanding of this regime by taking the $n\to \infty$ limit from the general formula of moments~\eqref{eq:momfinNsum}. In this limit, the moments are dominated by the largest $w_\alpha \sim \bar{w}_{e}$ near the edge. We can determine their value expanding $\Gamma(n+m-r)/
\Gamma(n-r) \simeq n^m$ in 
Equations~\eqref{eq:schurL} and \eqref{eq:momfinN} and performing the sum,
obtaining 
\begin{equation} 
\label{eq:Muni}
 \mathfrak{M}(m) := \lim_{n \to \infty} e^{-m \bar w_{e}} M(m) = \frac{2^{m-1}\sinh(2 m \gamma t )^{m-1}}{m!} \,,
\end{equation}
where we also fixed $\bar{w}_{e} = 2 n \gamma t + \log n$. This formula suggests the existence of non-trivial statistics governing the fluctuations of the Lyapunov exponents $w$'s at the edge. However, we note that it does not define a normalized distribution since $\mathfrak{M}(m \to 0) = \infty$. This is due to a crossover at small $m$ between a regime dominated by the bulk with $M(m \to 0) = n$ and one dominated by the edge $M(m) \stackrel{n\gg 1}{=}O(\bar{w}_e^m)$. An identical formula can be obtained in the analogous long-time limit considering the product of Wishart random matrices~(see Equation (3.41) in~\cite{Akemann_2013}). This is a manifestation of universality expected at large times in monitored systems and more generally in products of random matrices~\cite{deluca2023universality}.

The calculation of the moments of the trace in this regime is much more involved, but it can be expressed as in Equation~\eqref{eq:OmSchur}.
Indeed, in the limit of large $n$, we can replace $s_\kappa(1) \to n^m \chi_{1^m}^\kappa /m!$~\cite{macdonald1998symmetric}. One can understand this using $s_\kappa(y_1 = 1,\ldots, y_\hilbn = 1)$ counts the number of semistandard Young tableau of shape $\kappa$ and involving $\hilbn$ entries, whereas $\chi_{1^m}^\kappa$ counts the number of standard Young tableau with entries $1,\ldots, m$. At large $n$, the difference between semistandard and standard Young tableau becomes irrelevant and $s_\kappa(1) / \chi_{1^m}^\kappa \sim \binom{n}{m} \sim n^m /m!$. We thus obtain
\begin{equation}
\label{eq:Ouni}
\mathfrak{O}(m) := \lim_{n\to\infty} e^{-m \bar{w}_e} \Omega(m) = \frac 1 {m!}\sum_{\kappa \vdash m}  e^{4 \gamma t \nu(\kappa)} (\chi_{1^m}^\kappa)^2
\end{equation}
which was obtained in~\cite{deluca2023universality} in a different model as an additional manifestation of the universality.

\subsubsection{Asymptotics at Large $\gamma t$ \label{sec:asylarget}}

When $\gamma t \gg 1$, the separations are $|w_\al-w_\beta| \gg 1$ $\forall\al,\beta$. Additionally, their fluctuations are much smaller than their separation. In this regime, we can approximate in the potential \\
\begin{equation}
    \log \sinh |\frac{w_\alpha - w_\beta}{2}| \sim |\frac{w_\alpha - w_\beta}{2}| \,,
\end{equation}
which is the $1$-dimensional Coulombic repulsion. This kind of potential was recently studied in~\cite{flack2023out} in the context of ranked diffusion. This potential induces a force $f_\al = \frac 1 2 \sum_{\beta\neq\al} \mathrm{sign}(w_\al - w_\beta)$, which simply counts the number of slower particles in the back of $w_\al$, minus the faster ones in its front. Assuming a given ordering $w_1 \ll w_2 \ll \ldots \ll w_n$, we obtain the simple dynamics  
\begin{equation}
\label{eq:dw_asymptotic}
    dw_\al \sim 4\gamma v_\al dt + \sqrt{4\gamma} dB_\al  \,.
\end{equation}
where the drift velocity $v_\al := \frac 1 2 ( 2\al -n-1)$. 
This equation is an example of ranked diffusion (RD), namely, a diffusion process in one dimension where the $n$ particles undergo a drift term proportional to their respective rank \cite{ledoussal2022ranked,flack2023out}. In the RD problem, however, the particles can cross freely, whereas here
they cannot cross, which leads to different types of fluctuations at finite $t$. Nevertheless, in both models in the regime
$\gamma t \gg 1$, the 
equations decouple and can be solved separately, namely
\begin{equation} 
\label{eq:w_asymptotic}
    w_\al (t) \sim 4\gamma v_\al t + \sqrt{4 \gamma} B_\al(t)\,,
\end{equation}
where we recall that the $B_\al(t)$'s are $n$ independent standard Brownian motions,
each of variance $t$ at time $t$. Equation~\eqref{eq:w_asymptotic} gives some characterization of the joint probability distribution of the Lyapunov exponents $w$ in this large time $\gamma t \gg 1$ regime. It is not complete however, as there are $O(1)$ contributions of the
joint cumulants of the $w_\alpha$ which persist at infinite time. These were computed in 
Section III-D of Ref. \cite{flack2023out} by a saddle point method, and that calculation is easily extended to
the present model in Appendix \ref{app:saddle}.
As a result, Equation \eqref{eq:w_asymptotic} must be treated with care 
when computing exponential moments.

\section{Entanglement Entropies for Continuous Monitoring}

Now, we make use of the results obtained in the previous section for the unbiased ensemble $\langle \ldots \rangle_0$ to characterize the behavior of the entanglement entropies.

\subsection{Short Time Regime}

First of all, let us consider the Coulomb gas regime. In this case, we can use the fact that the moments are self-averaging, i.e., $n^{-1}\tr \tilde \rho^k \stackrel{\mbox{in law}}{\to}  \mu(k)$. Thus, from Equation~\eqref{eq:renyitrho},
we obtain
\begin{equation}
    \langle S_k \rangle =
    S_{\rm max} + \frac{1}{1-k}\log \Bigl[\frac{\mu(k)}{\mu(1)^k}\Bigr] = \log n + \frac{1}{1-k}\log\left[\frac{{L_{k-1}^{(1)}}(-k \tau )}{k}\right] \,,
\end{equation}
where the Von Neumann entropy can be recovered in the limit $k \to 1$. Using ${L_{k-1}^{(1)}}(-k \tau )/k = 1 + k(k-1) \tau /2 + O(\tau^2)$ for $\tau \ll 1$, and ${L_{k-1}^{(1)}}(-k \tau )/k = k^{k-2} \tau^{k-1}/(k-1)! + O(\tau^{k-2})$ for $\tau \gg 1$, we also obtain the asymptotic expansions
\begin{equation}
    \langle S_k \rangle =
    \begin{cases}
        S_{\rm max} - \frac{k \tau}{2} + 
        O(\tau^2) \,, & \tau \ll 1 \\
        \log \frac{n}{\tau} + \frac{\log k!}{k-1}-\log (k) \,, & \tau \gg 1
    \end{cases} 
\end{equation}
from which we can extract the $k\to 1$ limit
\begin{equation}
\label{eq:S1asytau}
    \langle S_1 \rangle =
    \begin{cases}
        S_{\rm max} - \frac{\tau}{2} + 
        O(\tau^2) \,, & \tau \ll 1 \\
        \log \frac{n}{\tau} -\gamma_E +1 \,, & \tau \gg 1
    \end{cases} \,,
\end{equation}
where $\gamma_E$ is the Euler--Mascheroni constant (not to be confused with the rescaled rate $\gamma$). 
Note that in this regime, the Born rule~\eqref{eq:BR} and unbiased outcomes~\eqref{eq:forcemeas} give the same result, since the weight of each trajectory factorizes from the rest. 

\subsection{Universal Regime}

As we discussed in Section~\ref{sec:uni}, this regime is much harder to address as the $w$'s are strongly correlated but the moments $M(m)$ and $\Omega(m)$ are not independent: they are dominated by the behavior around the edge (see Equations~(\ref{eq:Muni}) and (\ref{eq:Ouni})). However, we can still access the Von Neumann entropy by making use of Equation~\eqref{eq:VNrepl}. Equation~\eqref{eq:Muni} admits a simple analytic continuation for $m \to 1$ and we obtain
\begin{equation}
\label{eq:Muniexp}
    \mathfrak{M}'(1) = \log (2\sinh (2 \gamma  t))+\gamma -1 \,. 
\end{equation}
Unfortunately, the dependence on $m$ in Equation~\eqref{eq:Ouni} is much less transparent. A way to perform the analytic continuation was developed in~\cite{deluca2023universality}. Here, we analyze the asymptotic behavior. At small $\gamma t$'s,  because of the symmetry $\nu(\kappa') = \nu(-\kappa) $, where $\kappa'$ is the integer partition dual to $\kappa$, it is clear that $\Omega(m)$ is an even function of time for every $m$ . Thus, at small $\gamma t \ll 1$, we can conclude that $\Omega(m) = O(\gamma t)^2$. Up to this order, the dynamics of Von Neumann entanglement is fully captured by~Equation~\eqref{eq:Muniexp}. Thus, we have
\begin{equation}
    \langle S_1 \rangle = -\mathfrak{M}'(1) + O(t \gamma)^2 = -\log (4 \gamma  t)-\gamma_E+1+O(t\gamma)^2
\end{equation}
which connects nicely with the behavior for $\tau\gg1$ obtained in Equation~\eqref{eq:S1asytau}, recalling  that $\tau = 4 \gamma n t$.

\begin{figure}
    \centering
    \includegraphics[width=0.65\textwidth]{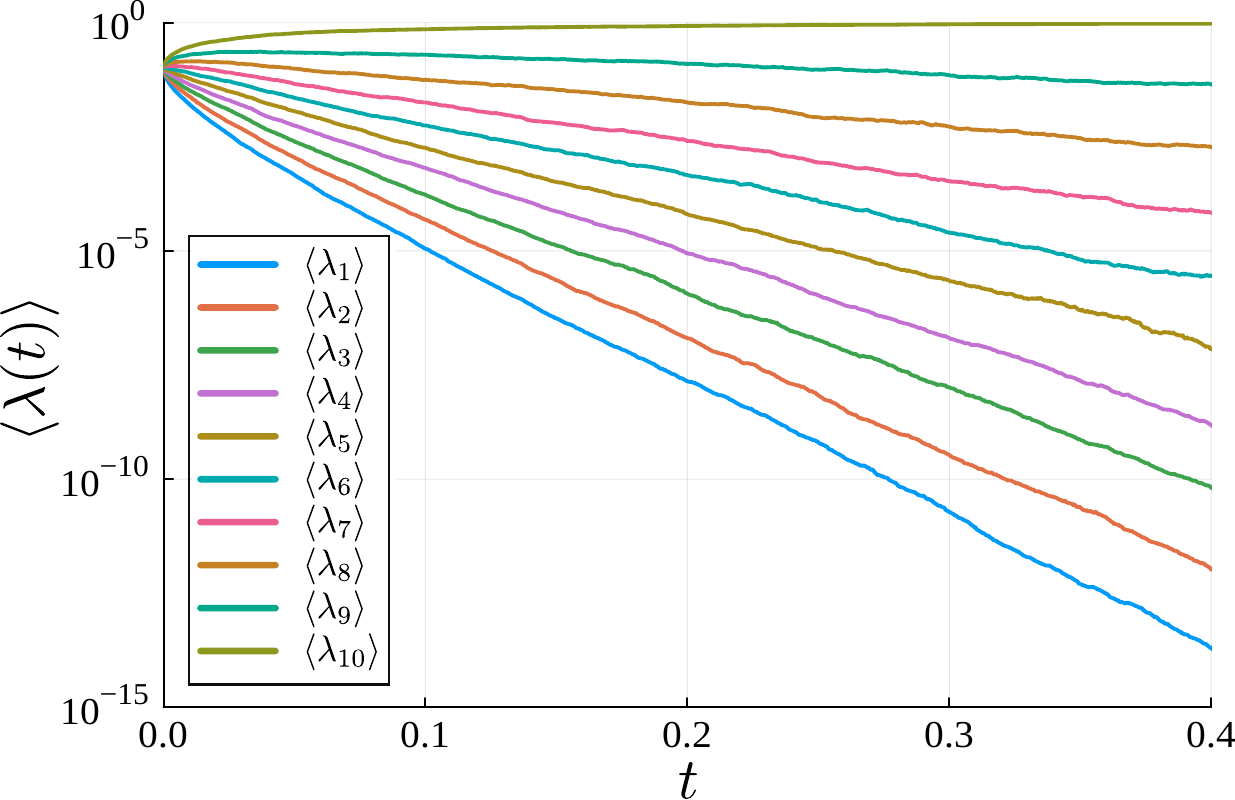}
    \caption{Long-time ranked diffusion. Long-time dynamics of the averaged eigenvalues $\lambda(t)$ for $x=0$, $\gamma=1$, $n=10$, whereas the maximum eigenvalue tends to one as $\braket{\lambda_{10}(t)}\sim 1-e^{-4 t}$ as time increases, the first $n-1$ eigenvalues are equispaced in logarithmic scale, i.e., $\braket{\lambda_\al(t)}/\braket{\lambda_\beta(t)}=e^{4(\alpha-\beta)t}$.}
    \label{fig:crystal}
\end{figure}

Although the regime of intermediate times is hard  to access, we can still estimate the asymptotic expansion $\gamma t \gg 1$. As discussed in Section~\ref{sec:asylarget}, the Lyapunov exponents separate linearly in time because of the drift Equation~\eqref{eq:w_asymptotic}. 
Returning to the eigenvalues $\lambda_\al = e^{w_\al} / \sum_\beta e^{w_\beta}$, it is evident that if $w_n \gg w_{\beta\neq n}$, then $\lambda_\al \sim e^{w_\al - w_n} \sim \delta_{\alpha,n}$, and the first $n-1$ eigenvalues are suppressed by purification. 
In particular, 
using Equation~\eqref{eq:w_asymptotic}, we also have that 
\begin{equation} \label{eq:ave_lambda_approx}
    \langle\lambda_{\al\neq n}(t)\rangle
    = \frac{\langle e^{w_\alpha} \rangle_0}{\langle \sum_\beta e^{w_\beta} \rangle_0} \sim \frac{\langle e^{4 \gamma t v_\alpha + \sqrt{4 \gamma} B_\alpha(t)} \rangle_B}
    {\langle e^{4 \gamma t v_n + \sqrt{4 \gamma} B_n(t))  } \rangle_B}
    \sim e^{(\al - n) 4\gamma t}\,,
\end{equation}
where $v_\alpha$ are the drift velocities in Equation~\eqref{eq:dw_asymptotic} and $\langle \dots \rangle_B$ denotes the average over the independent Brownians.
The sum of the first $n-1$ averages gives $\sum_{\al < n} \braket{\lambda_{\al}(t)} \sim e^{-4\gamma t}$, yielding 
\begin{equation}
\label{eq:maxlambda}
    1 - \braket{\lambda_n(t)} \sim e^{-4 \gamma t} \,.
\end{equation}
Moreover, the ratio between any two eigenvalues $\braket{\lambda_{\al\neq n}(t)}$, $\braket{\lambda_{\beta\neq n}(t)}$
\begin{equation}
    \frac{\braket{\lambda_\al (t)}}{\braket{\lambda_\beta (t)}} \sim e^{(\al - \beta) 4\gamma t} \,,
\end{equation}
which is independent of number $n$ of diffusing particles, and indicates that the first $n-1$ eigenvalues are logarithmically equispaced, as we show in Figure~\ref{fig:crystal}. 
Because of the exponential separation of the eigenvalues, from Equation~\eqref{eq:maxlambda}, we obtain the estimate for the Von Neumann entropy
\begin{equation}
    \begin{split}
        \langle S_1 \rangle & \sim - \frac{\langle e^{w_n} \log( e^{w_n}/\sum e^{w_\alpha} ) \rangle_0}{\langle e^{w_n} \rangle_0}= \frac{\langle e^{w_n} \log( 1 + \sum_{\alpha<n} e^{w_\alpha-w_n} ) \rangle_0}{\langle e^{w_n} \rangle_0} \approx 
        \frac{\langle e^{w_{n-1}} \rangle_0}{\langle e^{w_n} \rangle_0} = O(e^{-4 \gamma t}) \,,
    \end{split}
\end{equation}
which is in agreement with the $n=2$ result of Equation\,\eqref{eq:VN_t_n=2} if one replaces $\Gamma = 2 \gamma$. A more detailed discussion of this approximation is given in Appendix~\ref{app:saddle}.

Although this suggests that for any $n$ only the two largest eigenvalues matter for the calculations, the prefactor of the asymptotic expression of $\langle S_1 \rangle$ does \emph{not} reduce to the one from the $n=2$ model. Indeed, a more careful calculation (see Appendix\,\ref{app:saddleS}) shows that the saddle-point evaluation of the numerator of $\langle S_1 \rangle$ is shifted with respect to the one of the normalization, so that the two prefactors do not cancel out. In particular, we have 
\begin{equation} \label{eq:EEcrystal}
     \langle S_1 \rangle = \frac{n-1}{n}\, e^{-4 \gamma t} \left( 1 + \frac{1}{\sqrt{4 \pi \gamma t}} + O(t^{-1}) \right) \,, 
\end{equation}
which once again agrees with the result of the $n=2$ case, whereas the prefactor $(n-1)/n$ tends to 1 in the $n\to\infty$ limit.

\section{Conclusions}

In this work, we studied the purification dynamics induced by weak measurement, where the measurement operators are random matrices drawn from the GUE. The ensuing 
is written in terms of a stochastic Schr\"odinger equation, invariant under unitary transformation. 
Because of rotational invariance, we were able to determine the evolution of the density matrix eigenvalues $\{\lambda_\al\}_{\al=1}^n$, both for imperfect and perfect monitoring. 
In both cases, no MIPT takes place, as the system is far away from the critical point and is characterized by volume-law scaling of the entanglement entropy. 

In the former case of imperfect measurements $x>0$, we were able to thoroughly characterize the stationary distribution of the density matrix in the limit $n\to \infty$. There, the stationary joint probability distribution of the eigenvalues of the density matrix takes the typical expression of matrices from the inverse-Wishart ensemble, and the spectral density attains the corresponding inverse-Marchenko--Pastur form. Anyway, we showed that a solution for the finite-time stochastic dynamics also exists via a mapping to a quantum problem in imaginary time. In this context, the eigenvalue evolution at $x>0$ and for finite $n$ is still unknown. The relevant $n=2$ case seems to suggest that the stationary state does not belong to the inverse-Wishart ensemble, and that a more sophisticated stratagem must be sought in order to treat the non-conservative forces due to additional dephasing. 

In the case of perfect measurements $x=0$, we were able to fully determine the joint distribution function for the eigenvalues, at both finite time and finite $n$. Again, this can be conducted by mapping the stochastic dynamics onto fermions in imaginary time. Surprisingly, those undergo free diffusion as the external potential due to weak measurement disappears. Moreover, this allows us to factorize the joint probability distribution with respect to the $\tr \rho=1$ constraint, thus making the computation of average quantities accessible. We identify two different regimes. The first is a short-time one, where a Coulomb gas distribution appears with a time-dependent continuous density characterizing the spectrum of Lyapunov exponents. The density exhibits a characteristic crossover between the GUE semicircle and the uniform distribution. At later times, the Lyapunov exponents separate further and a continuous description is not possible anymore. Our calculations are perfectly consistent with the prediction of~\cite{deluca2023universality} and provide a new example supporting the universality of this regime. 
Subsequently, the Lyapunov exponents separate linearly over time, and the interactions between them become less important. In this regime, the dynamics resemble that of a ranked diffusion. 
We use this information to characterize the dynamics of entanglement in the various time regimes. 

Some comments are important. It would certainly be intriguing to characterize the crossover at $x\to0$ small and large times. Furthermore, from a technical point of view, it would be interesting to tie the regime emerging at long times with the spectral distribution of Lyapunov exponents near the edge.

Finally, interesting questions endure, characterizing the differences induced by studying the problem within other random matrix models, such as the Gaussian orthogonal ensemble.
\vspace{6pt}

\section{Acknowledgements}
We thank Alberto Rosso, Christophe Texier, and, in particular, Adam Nahum for discussions.
FG acknowledges support from Universit\'e Paris-Saclay. GG and ADL acknowledge support by the ANR JCJC grant ANR-21-CE47-0003 (TamEnt).
PLD acknowledges support from 
ANR grant ANR-23-CE30-0020-01 EDIPS,
and thanks KITP for hospitality, 
supported by NSF Grants No. NSF PHY-1748958 and PHY-2309135.

\paragraph*{Note added.} Recently, Ref.~\cite{bulchandani2024random} appeared, where a similar approach is independently studied in the case of unbiased outcomes. Equation~\eqref{eq:P0expr} is also obtained and the Coulomb gas regime is studied, also identifying the crossover between a GUE and a uniform distribution.

\appendix 

\section{Identities} \label{identities} 

To perform the calculation in the text we use the following two identities:
\begin{equation}
\begin{split}
\sum_{\alpha}\sum_{\beta\neq\alpha} 
 \coth \left( \frac{w_\alpha - w_\beta}{2}\right) \sum_{\gamma\neq\alpha,\beta}
 \coth\left( \frac{w_\beta - w_\gamma}{2}\right) 
 = \frac{n(n-1)(n-2)}{3} \,, 
\end{split}
\end{equation}
and 
\begin{equation}
\sum_\alpha 
 \sum_{\gamma \neq \alpha} \coth\left( \frac{w_\alpha - w_\gamma}{2} \right) \frac{e^{w_\alpha}}{\sum_\beta e^{w_\beta}} 
 = n-1 \,.
\end{equation}
The first one is well known from Calogero's papers (see references and generalizations in, e.g.,
Appendix A of \cite{10.21468/SciPostPhys.11.6.110}) 
and the second is specific to the present case.

\section{Finite-Time Von Neumann Entropy for the \boldmath{$n=2$} Case}
\label{finitetimeS}
We will now derive an expression for $\braket{S_1}(t)$ assuming $n=2$. In this case, the finite-time probability density is given by Equation\,\eqref{eq:distrib_n2}, whereas we have 
\begin{equation}
\begin{split}
    \braket{S_1} = \langle \frac{\omega e^{-\omega}}{\cosh \omega}   \rangle + \langle \ln (1 + e^{-2\omega}) \rangle = \langle \frac{\omega e^{-\omega}}{\cosh \omega}   \rangle - \sum_{p=1}^\infty \frac{(-1)^{p}}{p} \langle e^{-2 p \omega} \rangle \,. 
\end{split}
\end{equation}
Replacing the explicit form of $P(\omega,t)$ we obtain (setting $\Gamma =1$ for simplicity)
\begin{equation}
\begin{split}
    \braket{S_1} = & \frac{e^{-2t}}{2} \left( 1 + \sqrt{\frac{8t}{\pi}} \right) - \frac{1 + 4t}{2} \text{erf} (\sqrt{2t}) + \\
    & - \frac{\sqrt{2t}}{2} \sum^{\infty}_{p=1}(-1)^{p}e^{2p^2t} \Big[ e^{4pt} \left( 1 + \frac{1}{p} \right) \text{erf}(p+1) - e^{-4pt} \left( 1 - \frac{1}{p} \right) \text{erf} (p-1) \Big] \,. 
\end{split}
\end{equation}
Expanding for large times, and considering $\text{erf}(x) = e^{-x^2}/\sqrt{2 \pi x} \left( 1 - x^{-2}/2 + O(x^{-4}) \right)$, we have
\begin{equation}
\begin{split}
    \braket{S_1} = \frac{e^{-2t}}{2} \big[ 1 - \frac{t^{- \frac{3}{2}}}{2 \sqrt{2 \pi}} + \frac{t^{- \frac{1}{2}}}{\sqrt{2 \pi}} -\frac{t^{- \frac{3}{2}}}{16 \sqrt{2 \pi}} - \frac{t^{- \frac{3}{2}}}{\sqrt{2 \pi}} \sum^{\infty}_{n=2} \frac{(-1)^n}{(n^2-1)^2} + O(t^{-\frac{5}{2}}) \big] \, , 
\end{split}
\end{equation}
hence making use of the identity 
\begin{equation}
    \sum_{n=2}^{\infty}  \frac{(-1)^n}{(n^2-1)^2} = \frac{1}{4} \zeta(2) - \frac{5}{16} = \frac{\pi^2}{24} -  \frac{5}{16} \, 
\end{equation}
we find the result \eqref{eq:VN_t_n=2} present in the main text. 

\section{Inverse-Wishart Ensemble} \label{inverseWishart}
The unitary ($\beta=2$) Wishart ensemble is defined by the distribution 
\begin{equation} \label{eq:wishart_ensemble}
    P(A) = (\det A)^{m-n} e^{-1/2 \tr A} \,,
\end{equation}
for any squared $n\times n$ matrix $A$, with integer $m\geq n$ \cite{livan2018introduction, potters2020first}. 
A Wishart-distributed matrix $A$ can be written as $A=HH^\dagger$, with $H$ a rectangular $n\times m$ matrix with complex Gaussian entries. 
Equation~\eqref{eq:wishart_ensemble} leads to the following joint distribution for the eigenvalues $\{a_i\}_{i=1}^n$, as follows: 
\begin{equation}
    P(\vec a) = \prod_{i>j} (a_i - a_j)^2 \prod_j a_j^{m-n} e^{-1/2 \sum_j a_j} \,.
\end{equation}
The marginal distribution of the eigenvalues, i.e., the spectral density, can be obtained from the previous equation. In the limit of infinitely large matrices $n\to\infty$, a standard result is the Marchenko--Pastur distribution 
\begin{equation} \label{eq:f_MP}
    f_{\mathrm{MP}}(\tilde a) = \frac{1}{2\pi a} \sqrt{(\tilde a-\tilde a_-)(\tilde a_+-\tilde a)} \,,
\end{equation}
with the rescaled eigenvalues $\tilde a = a/2n$. The endpoints of $f_{\mathrm{MP}}$ are defined by $\tilde a_\pm = (1\ \pm\sqrt{m/n})^2$.  

It is then possible to derive the inverse-Wishart ensemble of matrices $B = A^{-1}$. In terms of the eigenvalues $b_i = a_i^{-1}$, one simply has to find the distribution as follows: 
\be \begin{split}
P(\vec b)  = \prod_j \frac{da_j}{db_j} P(\vec a) =  \prod_j b_j^{-m-n} e^{- \frac{1}{2} \sum_j b_j^{-1}} \prod_{i>j} (b_i - b_j)^2 \,,
\end{split}
\ee
or, in matrix terms 
\begin{equation}
    P(B) = (\det B)^{-m-n} e^{-1/2 \tr B^{-1}} \,.
\end{equation}
So it is easy to see that for $n=2$, we have $m=1$ which does not fulfill the condition $m\geq n$.
The inverse-Marchenko--Pastur distribution for the eigenvalues of an inverse-Wishart matrix simply reads, following from Equation~\eqref{eq:f_MP}:
\begin{equation} \label{eq:f_IMP}
\begin{split}
    f_{\mathrm{IMP}}(\tilde b) = \frac{d\tilde a}{d\tilde b} f_{\mathrm{MP}}(\tilde a) = \frac{m/n -1}{2\pi \tilde{b}^2} \sqrt{(\tilde b-\tilde b_-)(\tilde b_+-\tilde b)} \,,
\end{split}
\end{equation}
with the inverse rescaled eigenvalues $\tilde b = \tilde a^{-1} = 2n b$. The new endpoints are $\tilde b_\pm = 1/\tilde a_\mp$. 
Clearly, the spectral density in Equation~\eqref{eq:f_IMP} is not well defined in the limit $m=n$. 

\section{Equivalent Dyson Brownian Motion} \label{DBM} 
Consider the DBM $\vec x(s)$ for $\beta=2$ 
\bea \label{dbm1}
dx_i(s) = \frac{1}{n}  \sum_{j \neq i} \frac{1}{x_i-x_j} ds
+ \frac{1}{\sqrt{n}} db_i(s) \,,
\eea 
where the $db_i(s)$ are independent standard Brownian motions, with $db_i(s) db_j(s) =\delta_{ij} ds$,
and we choose a fixed ordered initial condition $x^0_1>x^0_2>\dots>x^0_N$. 
At fixed time $s$, $\vec x(s)$ has the same law as the spectrum of the random matrix $X= {\rm diag}(x_i^0) + \sqrt{s} H$. 
Its propagator takes the form
\be 
P_{DBM}(\vec x , s  | \vec x^0, 0 )
=  \frac{\Delta(\vec x)}{\Delta(\vec x^0)}  \det G_0\left(x_i,x^0_j,\frac s n \right)  \,.
\ee 
Until now, the initial condition is arbitrary. We now choose $x^0_j=\delta_j/n$, 
regularly spaced. In this case, $\det G_0(x_i,x^0_j,s/n)$
can be explicitly evaluated and one finds
for $x_1>x_2>\dots>x_N$ as follows:
\begin{equation}
    \begin{split}
        P_{DBM}(\vec x , s  | \vec \delta,0 ) \frac{1}{Z^{\rm DBM}_s} \Delta(\vec x) \, \prod_{j>i}  \sinh(\frac{x_i-x_j}{2s}) \, e^{- \frac{n}{2 s} \sum_i x_i^2 }  \,.
    \end{split}
\end{equation}
To make the connection with the distribution $P_0(\vec w,t)$ of the main text, we set $s=1/(4\gamma t n)$ and $x_j=w_j/(4 \gamma t n)$. Then, one recovers $P_0$; namely, one has
\be 
\frac{1}{(4 \gamma t n)^{n}} \, P_{DBM}\Big(\left.\frac{\vec w}{4 \gamma t n} , \frac{1}{4 \gamma t n}  \right| \frac{\vec \delta}{n},0 \Big) = P_0(\vec w,t) \,.
\ee 
This is in agreement with \eqref{X} in the main text. 
The above correspondence is exact and valid for any $n$. We now consider the large $n$ limit. 
In the text, we scaled time as $t=\frac{\tau}{4 \gamma \hilbn}$ and considered the limit $n \to \infty$ at fixed $\tau$. As mentioned in the text around \eqref{X}, we can focus
on the DBM $\vec x(s)$ in inverse time $s=1/\tau$. 
Let us denote $\mu_s(x)$ as its density, and $h(z,s)= \frac{1}{N} \sum_i \frac{1}{z-x_i(s)}$ its resolvent.
We will now compute both quantities, and from them we will deduce 
\be \label{rescg} 
\rho_\tau(w)  = \frac{1}{\tau} \mu_{1/\tau}(\frac{w}{\tau})
\,, \quad 
g(z,\tau) = \frac{1}{\tau} h(\frac{z}{\tau},\frac{1}{\tau}) \,,
\ee 
which are displayed in the text. The DBM density $\mu_s(x)$ interpolates between being uniform in $[-1/2,1/2]$ at small $s$, and a semi-circle of support $\sqrt{s} [-2,2]$ at large $s$. To compute it at all times, we recall that its resolvent $h(z,s)$ satisfies in the large $n$ limit the complex Burgers equation 
\be 
\partial_s h = -  \frac{1}{2} \partial_z h^2  \,.
\ee 
The general solution is 
\be 
h(z,s)=h_0(u) \,, \quad z = u + s h_0(u) \,.
\ee 
Here, we have a uniform initial density
\begin{equation}
    h_0(u) = \int_{-1/2}^{1/2} \frac{dx}{u-x} = \log \frac{u+\frac{1}{2}}{u-\frac{1}{2}}  \,,  \quad u = \frac 1 2 \coth(\frac{h_0}{2}) \,.
\end{equation}
Hence, one finds the parametric solution 
\be \label{zsolu}
z = s h + \frac 1 2 \coth(\frac{h}{2}) \,. 
\ee 
Upon rescaling \eqref{rescg} one obtains \eqref{zg} in the text. To compute the DBM eigenvalue density $\mu_s(x)$, one sets $z=x-i 0^+$ and $h=h_r + i \pi \mu$. Taking the imaginary part of \mbox{\eqref{zsolu} gives}
\be 
\cosh h_r = a_s(\mu) := \cos \pi \mu + \frac{\sin \pi \mu}{2 s \pi \mu} \,.
\ee 
{Inserting} $h_r$ within the real part of \eqref{zsolu}, one finds
\begin{equation} \label{densmu}
\begin{split}
 x&  = \pm s +\bigg[\frac{\pi \mu}{\sin (\pi \mu)} \sqrt{a_s(\mu)^2-1} +  \log \bigg(  a_s(\mu) + \sqrt{a_s(\mu)^2-1} \bigg) \bigg] \,,
\end{split}
\end{equation}
where the two branches correspond to $h_r>0$ and $h_r<0$. Equation \eqref{densmu} determines $\mu_s(x)$ for a given $s$. Upon 
the rescaling \eqref{rescg}, one obtains \eqref{eq:densrho} in the text. Note that an analogous calculation
was performed in \cite{Mergny_2021} in a different context. 

\section{Kernel \label{app:kernel}}

From the relation to the DBM described in the previous Appendix, and from Ref.~\cite{johansson2001universality}, one knows that the $\vec w$ form a determinantal point process. This means that both their joint PDF and their correlation functions (obtained by integrating over some of the $w_\alpha$) are equal to determinants involving a kernel. Here, we obtain a convenient form for this kernel using bi-orthogonal polynomials. We follow the method of Ref.~\cite{Akemann_2013}, c.f. Equations (3.11)--(3.13) therein.
To simplify the expressions, in this section we fix $\gamma = 1/4$. Equivalently, one can recover the general form by replacing $t \to 4 \gamma t$.
One defines the two sets of polynomials
\begin{align}
Q_n(w;t) &:=   \sum _{l=0}^n \frac{(-1)^{n-l} \exp \left(-\frac{l^2 t}{2}-l w\right)}{\Gamma (l+1) \Gamma (-l+n+1)} \,, \\
P_n(w;t) &:= \sum_{i=0}^n \left(\frac{-1}{\sqrt{2 t}}\right)^i S_n^{(i)} H_i\left(\frac{w}{\sqrt{2 t}}\right) \,,
\end{align}
where $S_n^{(i)}$ is the Stirling number of the first kind. 
These polynomials are not monic, but they are normalized to have unit integral. Indeed, they satisfy the orthogonality relation
\begin{equation}
    \int \frac{dw}{\sqrt{2 \pi t}} e^{-\frac{w^2}{2t}} P_n(w;t) Q_{n'}(w;t) = \delta_{n,n'} \,.
\end{equation}
From these polynomials, one defines a first kernel as
\begin{equation}
    \tilde K_n(w, w';t) = \sum_{\ell=0}^{n-1} P_n(w;t) Q_n(w';t) \frac{e^{- \frac{w^2 + w'^2}{4 t} } }{\sqrt{2 \pi t} } \,.
\end{equation}
To obtain a centered process, however, one defines the shifted kernel
\be 
K_n(w, w';t) = \tilde K_n(w - t \frac{n-1}{2} , w' - t \frac{n-1}{2}  ; t) \,.
\ee 
Both kernels are self-reproducing, i.e., one has $K_n^2=K_n$. 
The density of the $\vec w$ for any $n$ and any time $t$ is then given by 
\be 
f_n(w,t) = \frac{1}{n} K_n(w,w;t) \,,
\ee 
and it is normalized to unity. 
One can check that one has indeed
\be 
P_0(\vec w,t) =  \frac{1}{n!}  \det_{1 \leq i,j \leq n} K_n(w_i,w_j;t) \,.
\ee 

We can also use a similarity transformation to re-write the Kernel as
\begin{equation}
    \tilde{K}_n(w, w') = J_n(w',w) e^{\frac{w'^2 - w^2}{4t}} \,,
\end{equation} 
with 
\begin{equation}
    \begin{split}
        J_n(w,w') = \sum_{k=1}^n e^{- t (w/t-y_k)^2/2} \, \int_{i \mathbb{R}} \frac{dz}{2 i \pi} e^{t \frac{(z-w'/t)^2}{2}} \prod_{j \neq k =1}^n \frac{z- y_j}{y_k - y_j}   \,,
    \end{split}
\end{equation}
with $y_k=1-k$. This last form is analogous to Proposition 2.3 and Equation (2.18) in \cite{johansson2001universality}. We can further expand the last term, using
\begin{equation}
\label{eq:gammaprod}
    \prod_{j \neq k=1}^n \frac{z- y_j}{y_k - y_j} = \frac{1}{z - (1-k)}\frac{\Gamma (n+z)}{\Gamma (z)} \frac{(-1)^{k-1}}{(k-1)! (n-k)!}\,.
\end{equation}
One can check that the exact expression for the moments $M(m)$ in \eqref{eq:momfinNsum} are recovered 
from $M(m)$ $=$ $e^{-m(n-1)t/2}$ $\int dw J_n(w, w) e^{w m}$ (since the quadratic term in $w$ cancels with the one in $w'$,
the integration over $w$ leads to a delta function), where the prefactor is related to the shift with respect to the edge position.

This form can be used to derive the limit of the kernel when $n \to \infty$, focusing on the neighborhood of the edge. We set
\begin{equation}
    J_\infty(\omega, \omega') = \lim_{n\to\infty} \, e^{\frac{\omega - \omega'}{t})\log n } J_n(\omega + \log n, \omega' + \log n) \,,
\end{equation}
which is defined so that the largest Lyapunov is moved at $\omega = 0$. From the explicit limit, we obtain the expression
\begin{equation}
J_\infty(\omega,\omega')  =  
 \sum_{k=1}^\infty  \frac{(-1)^{k-1}}{(k-1)!} \int_{i \mathbb{R}} \frac{dz}{2 i \pi}  
e^{-\frac{(\omega -(1-k) t)^2}{2 t} + \frac{(\omega'-t z)^2}{2 t}} \frac{1}{(z - (1-k))\Gamma(z)}  \,.
\end{equation}
Exponentiating the denominator with an integral, we can also express
\begin{equation}
    J_\infty(\omega,\omega') = e^{ \frac{\omega^{\prime 2} - \omega^{2}}{2 t}} \int_{\mathbb{R}^+} dr A(\omega+r) B(\omega'+r) \,,
\end{equation}
with
\begin{equation}
        A(\omega) = \sum_{k=1}^\infty \frac{(-1)^{k-1}}{(k-1)!} e^{-t \frac{(1-k)^2}{2} + \omega (1 - k) } \,, \quad 
        B(\omega') = \int_{i \mathbb{R}} \frac{dz}{2 i \pi} e^{t \frac{z^2}{2} - z \omega' } \frac{1}{\Gamma(z)} \,.
\end{equation}
This is the kernel describing the Lyapunov exponents in the universal regime $\Gamma t = O(n)$. 
A similar kernel was obtained in formula (1.15) in \cite{Liu2023}
and in formula VI.22 in \cite{PhysRevE.102.052134} in the context of the
universal edge statistics of products of Ginibre matrices. The connection
to the Dyson Brownian motion was also discussed in \cite{PhysRevE.102.052134}.

\section{Large Time Moments from a Saddle Point}
\label{app:saddle} 
We perform here a calculation analogous to the one in Section III-D of Ref.~\cite{flack2023out}.
It is valid for any $n$.
Let us denote $\mathcal{O}$ the ordered sector $w_1<\dots<w_n$. For $\vec w \in \Omega$, one can rewrite
exactly (not keeping track of time-dependent normalizing constants) $P_0(\vec w,t) \sim e^{-S}$, with $S=S_0+S_{\rm int}$ as follows:
\begin{equation} \label{S}
        S_0 = \sum_\alpha \frac{(w_\alpha - 4 \gamma t v_\alpha)^2 }{8 \gamma t} \,, \quad 
        S_{\rm int} = - \sum_{\alpha<\beta} \Big[ \log|w_\beta-w_\alpha| + \log\left(1 - e^{- (w_\beta-w_\alpha)}\right) \Big] \,,
\end{equation}
recalling that $v_\alpha=\alpha - \frac{n+1}{2}$. Let us compute $G[\vec m]= \langle e^{\sum_\alpha m_\alpha w_\alpha} \rangle_0$.
Changing variables to $w_\alpha=4 \gamma t z_\alpha$, one finds
\begin{equation}
    G[\vec m] \sim \int_{\vec z \in \Omega} e^{- 4 \gamma t \tilde S} e^{- S_{\rm int}} \,, \quad \tilde S=\sum_\alpha \frac{1}{2} (z_\alpha - v_\alpha)^2 - \sum_\alpha m_\alpha z_\alpha
\end{equation}
For $\gamma t \gg 1$ the term $e^{- 4 \gamma t \tilde S}$ has a saddle point for 
\be 
z_\alpha = z_\alpha^* = v_\alpha + m_\alpha \,.
\ee 
The saddle point remains in the ordered sector $\Omega$, providing $m_\alpha-m_{\alpha-1} +1 >0$ for all $\alpha$. That case corresponds to the particle crossing being irrelevant. The interaction term takes the form (up to a time-dependent constant)
\begin{equation}
        S_{\rm int} = - \sum_{\alpha<\beta} \Big[  \log|z_\beta-z_\alpha| + \log \left(1 - e^{- 4 \gamma t (z_\beta-z_\alpha)} \right) \Big] \,.
\end{equation}
For $\gamma t \gg 1$, the last term is irrelevant compared to the first (provided the particle crossings are irrelevant). The saddle-point method then leads to
\begin{equation}
        G[\vec m]  \simeq e^{ 4 \gamma t \sum_\alpha ( m_\alpha v_\alpha + \frac{1}{2} m_\alpha^2)} \prod_{\alpha<\beta} \frac{ \beta - \alpha + m_\beta - m_\alpha}{\beta - \alpha} \,.
\end{equation}
Taking derivatives of $\log G[\vec m]$ gives all the $O(1)$ joint cumulants of the variables $w_\alpha$ in the large time limit. This estimate is valid as long as the saddle point is in the $\mathcal{O}$ sector.

Specializing $m_\alpha=\delta_{\alpha, n}$, i.e., the largest of the $w$'s, gives 
\be \label{momn}
\langle e^{m w_n} \rangle_0 \simeq
\binom{m+n-1}{n-1} \, e^{2 \gamma t m (m+n-1) } \,.
\ee 
Since $\sum_\alpha e^{m w_\alpha}$ is dominated by the term $\alpha=n$, this result agrees with the formula \eqref{eq:momfinNsum} for $M(m)$ at large time (from which we see that corrections are $O(e^{- 4 \gamma t})$). For a general $\alpha$, one obtains instead
\begin{equation} \label{momalpha} 
        \langle e^{m w_\alpha} \rangle_0 \simeq e^{2 \gamma t m (m+2 \alpha - n -1) } \frac{\sin (\pi  m) \Gamma (m+\alpha) \Gamma (-m+n-\alpha +1)}{\pi m \Gamma (\alpha ) \Gamma (n-\alpha +1)}\,. 
\end{equation}
In these results, we see that the prefactor includes the interactions with all the particles (not just the neighbor). Furthermore, we see from the condition
that the saddle point remains in the $\Omega$ sector that 
(i) Equation \eqref{momn} is valid for all $m>-1$ (in agreement with \eqref{eq:momfinNsum}) 
(ii) Equation \eqref{momalpha} for $\alpha<n$ is valid only for $-1<m<1$. Indeed, for $m>0$, the rightmost particle is pulled out of the gas, and hence particle crossings are irrelevant. An ``internal" particle, however, cannot be pulled too strongly without crossing its neighbors. 

Finally, note that evaluating $\Omega(m)= \langle (\sum_\alpha e^{w_\alpha})^m \rangle_0$ by the same method leads to an additional term $- m z_n$ in $\tilde S$, plus another term
$- \log(1+ \sum_{\alpha<n} e^{- 4 \gamma t (z_\alpha-z_n)})$, irrelevant at large times. Hence, we find $\Omega(m) \simeq M(m)$ by this saddle-point method in the $\gamma t \gg 1$ limit.

\section{Long-Time Entanglement Entropy}
\label{app:saddleS}
Here, we derive the long-time asymptotic expression for the entanglement entropy, for arbitrary value of $n$. 
We start from the expression of the entanglement entropy as computed in Equation\,\eqref{eq:VNrepl}, namely
\begin{equation}
    \langle S_1 \rangle = \left.\partial_m \log \frac{\Omega(m)}{M(m)} \right|_{m=1}  \,,
\end{equation}
where $M(m)$ and $\Omega(m)$ are defined in Equations \eqref{eq:momMdef} and \eqref{eq:momtrdef}, respectively, and the average $\langle \cdot \rangle_0$ is taken on the probability distribution in Equation\,\eqref{eq:VNrepl}. Let us set $ 4 \gamma t = t$ for the rest of the section, and let us define $w_\alpha = u_\alpha t$, so that we have
\begin{align}
    & M(m) \propto\int_{<} d^{n}u \, e^{- t \sum_\alpha \frac{u_\alpha^2}{2}} \left( \sum_\alpha e^{m u_\alpha t}\right) \prod_{\alpha > \beta} \sinh \frac{t(u_\alpha - u_\beta)}{2} (u_\alpha-u_\beta)  \,,\\
    & \Omega(m) \propto \int_{<}  d^{n}u \, e^{- t \sum_\alpha \frac{u_\alpha^2}{2}}  \left( \sum_\alpha e^{u_\alpha t} \right)^m   \prod_{\alpha > \beta} \sinh \frac{t(u_\alpha - u_\beta)}{2} (u_\alpha-u_\beta) \,,
\end{align}
with the notation $ \int_{<} \equiv \int_{u_1 < u_2 < \dots < u_n}$. We exploited therein the permutation symmetry of the original integrals to order the new variables $u_\alpha$. For large $t$
\begin{align}
    & \ln \sinh \frac{t(u_\alpha - u_\beta)}{2} = 
    \frac{t}{2} (u_\alpha - u_\beta) - \ln 2 + e^{t(u_\beta - u_\alpha)}  \\
    & \ln  \sum_\alpha e^{m u_\alpha t}  = mt u_n  + e^{m t(u_{n-1}-u_n)} \\
    & \ln  \big(\sum_\alpha e^{u_\alpha t} \big)^m  = mt u_n  + m e^{t(u_{n-1}-u_{n})} 
\end{align}
up to exponentially small corrections in $t$, so that we have
\begin{align}
    & M(m) \propto\int_{<} d^n u \, \Delta(u) \, e^{- t (I(u)- m u_n)} \Bigg( 1 + e^{ - m (u_n -u_{n-1}) t} - \sum_{\alpha > \beta} e^{-t(u_\alpha - u_\beta)} \Bigg)  \;,\\
    & \Omega(m) \propto \int_{<} d^n u\, \Delta(u) \, 
e^{- t (I(u)- m u_n)} \Bigg( 1 + m e^{ - (u_n -u_{n-1}) t} - \sum_{\alpha > \beta} e^{-t(u_\alpha - u_\beta)} \Bigg) \,,
\end{align}
where $\Delta(u) \equiv \prod_{\alpha > \beta} (u_\alpha-u_\beta)$ is the Vandermonde determinant and
\begin{equation}
    I(u) = \sum_\alpha \frac{u_\alpha^2}{2} - \frac{1}{2} \sum_{\alpha > \beta} (u_\alpha - u_\beta) . 
\end{equation}
Finally, the expression of the entanglement entropy becomes
\begin{equation} \label{eq:appEEcrystal}
    \langle S_1 \rangle = \frac{ \int_{<} \, d^{n} u  \Delta(u)  \,  e^{- t (I(u)- u_{n-1})} [1 + t (u_{n}- u_{n-1})] }{\int_{< } d^{n} u \, \Delta(u) \,  
e^{- t (I(u)- u_n)}} \,, 
\end{equation}
which is exact up to exponentially small corrections. Because of the different exponential factors, the numerator and the denominator will have different saddle points. Let us now perform the calculations separately.
\\\\
\emph{Numerator: 
}  imposing $\partial_{u_\alpha} (I(u)- u_{n-1})|_{v_\alpha} = 0$, one finds the solutions
\begin{equation}
    v_\alpha = \frac{1}{2} (2 \alpha - 1- n + 2 \delta_{\alpha, n-1}) \,.
\end{equation}
Let us notice that $v_n = v_{n-1} > v_{n-2}> \dots > v_1$. It is now convenient to introduce new variables $u_\alpha = v_\alpha + \xi_\alpha/\sqrt{t}$. The integration domain is now 
\begin{equation}
    \xi_{\alpha-1} - \xi_{\alpha} < \sqrt{t} (v_\alpha - v_{\alpha-1}) \,,
\end{equation}
so that in the limit $t \rightarrow \infty$, we only have the constraint $\xi_{n-1}  < \xi_{n}$, up to exponentially small corrections. Let us now consider the Vandermonde determinant. The pair $(n-1,n)$ yields a factor $t^{-1/2} (\xi_n - \xi_{n-1})$. Instead, the other terms give the following: 
\begin{equation}
\begin{split}
    \Delta(u) = \tilde{\Delta} (v) \left( 1 + t^{-1/2} \sum_{\substack{\alpha>\beta \\ (\alpha, \beta) \neq (n,n-1)}} \frac{\xi_\alpha - \xi_\beta}{v_\alpha - v_\beta} \right) \,,
\end{split} 
\end{equation}
up to $O(t^{-1})$ terms, where we denoted with the symbol $\sum_{\alpha>\beta}^*$ the sum over couples such that $\alpha>\beta$ and $(\alpha, \beta) \neq (n,n-1)$, and with $\tilde{\Delta} (u)$ the Vandermonde determinant restricted to all pairs but the $(n,n-1)$ one. Let us notice that, by parity, the $O(t^{-1/2})$ terms vanish once integrated on the Gaussian measure. Finally, we have that, up to $O(t^{-1})$ terms, the numerator becomes 
\begin{equation}
\begin{split}
&t^{-n/2} e^{-t(I(v) - v_{n-1}) } \tilde{\Delta} (v) \int_{ < } d^n \xi e^{-\sum_{\alpha} \xi^2_{\alpha}/2}  \left[ (\xi_n - \xi_{n-1})^2 + t^{-1/2} (\xi_n - \xi_{n-1}) \right] = \\
&= (2 \pi)^{n/2-1} t^{-n/2} e^{-t(I(v) - v_{n-1}) }  \tilde{\Delta}(v) \cdot\\
&\qquad \cdot\int_{\xi_{n-1} < \xi_n} d\xi_n d \xi_{n-1} \, e^{-(\xi^2_{n} + \xi^2_{n-1})/2} \left( (\xi_n - \xi_{n-1})^2+ t^{-1/2} (\xi_n - \xi_{n-1}) \right) \,. 
\end{split}
\end{equation}
Introducing the rotated variables $\xi_{\pm} = (\xi_n \pm \xi_{n-1})/\sqrt{2}$ the last integral becomes trivial, so we have the following for the numerator:
\begin{equation} \label{eq:appnumerator}
(2 \pi)^{n/2} t^{-n/2} e^{-t(I(v) - v_{n-1}) } \tilde{\Delta}(v)  (1 + (\pi t)^{-1/2}) \,.
\end{equation}
\emph{Denominator:} imposing again the condition $\partial_{u_\alpha} (I(u)- u_{n})|_{v^*_\alpha} = 0$, one finds the solutions
\begin{equation}
    v^*_\alpha = \frac{1}{2} (2 \alpha - 1- n + 2 \delta_{\alpha, n}) \,, 
\end{equation}
with instead $v^*_n > v^*_{n-1} > \dots > v^*_1$. In terms of the new variables $u_\alpha = v^*_\alpha + \xi_\alpha/\sqrt{t}$ the integration domain is now 
\begin{equation}
    \xi_{\alpha-1} - \xi_{\alpha} < \sqrt{t} (v^*_\alpha - v^*_{\alpha-1}) \,,
\end{equation}
so that in the limit $t \rightarrow \infty$ the integral will run on the whole $\mathbb{R}^n$ up to exponentially small correction. In this case, the $(n,n-1)$ pair in the Vandermonde determinant should not be singled out. Analogously to the numerator case, moreover, the $O(t^{-1/2})$ corrections coming from {the} Vandermonde determinant vanish once integrated. Finally, for the denominator we have the following: 
\begin{equation} \label{eq:appdenominator}
    (2 \pi)^{n/2} t^{-n/2} e^{- t (I(v^{*})- v^{*}_n)} \Delta(v^{*})  \,, 
\end{equation}
up to $O(t^{-1})$ corrections.
\\\\
\emph{Entanglement Entropy:} replacing the results \eqref{eq:appnumerator} and \eqref{eq:appdenominator} in the expression \eqref{eq:appEEcrystal}, we find
\begin{equation}
    \langle S_1 \rangle = \frac{\tilde{\Delta}(v)}{ \Delta(v^*)} e^{- t (I(v) - I(v^{*}) - v_{n-1} + v^{*}_n)} \left( 1 + (\pi t)^{-1/2} \right) \,.
\end{equation}
Taking into account the explicit form of $v_\alpha$ and $v^*_\alpha$, we have 
\begin{equation}
    I(v) - I(v^{*}) - v_{n-1} + v^{*}_n = 1 \,, 
\end{equation}
whereas the prefactor reads as follows:
\begin{equation}
   \frac{\tilde{\Delta}(v)}{ \Delta(v^*)} = \frac{\prod_{\alpha=1}^{n-2}  (n - \alpha) \prod_{\alpha=1}^{n-2}  (n - \alpha)}{ (v^{*}_n - v^{*}_{n-1}) \prod_{\alpha=1}^{n-2}  (n +1 - \alpha) \prod_{\alpha=1}^{n-2}  (n -1 - \alpha)} = \frac{n-1}{n} \,.
\end{equation} 
Restoring $t = 4\gamma t$, we reach the result of Equation~\eqref{eq:EEcrystal} of the main text.


\begin{thebibliography}{0}%
\makeatletter
\providecommand \@ifxundefined [1]{%
 \@ifx{#1\undefined}
}%
\providecommand \@ifnum [1]{%
 \ifnum #1\expandafter \@firstoftwo
 \else \expandafter \@secondoftwo
 \fi
}%
\providecommand \@ifx [1]{%
 \ifx #1\expandafter \@firstoftwo
 \else \expandafter \@secondoftwo
 \fi
}%
\providecommand \natexlab [1]{#1}%
\providecommand \enquote  [1]{``#1''}%
\providecommand \bibnamefont  [1]{#1}%
\providecommand \bibfnamefont [1]{#1}%
\providecommand \citenamefont [1]{#1}%
\providecommand \href@noop [0]{\@secondoftwo}%
\providecommand \href [0]{\begingroup \@sanitize@url \@href}%
\providecommand \@href[1]{\@@startlink{#1}\@@href}%
\providecommand \@@href[1]{\endgroup#1\@@endlink}%
\providecommand \@sanitize@url [0]{\catcode `\\12\catcode `\$12\catcode `\&12\catcode `\#12\catcode `\^12\catcode `\_12\catcode `\%12\relax}%
\providecommand \@@startlink[1]{}%
\providecommand \@@endlink[0]{}%
\providecommand \url  [0]{\begingroup\@sanitize@url \@url }%
\providecommand \@url [1]{\endgroup\@href {#1}{\urlprefix }}%
\providecommand \urlprefix  [0]{URL }%
\providecommand \Eprint [0]{\href }%
\providecommand \doibase [0]{https://doi.org/}%
\providecommand \selectlanguage [0]{\@gobble}%
\providecommand \bibinfo  [0]{\@secondoftwo}%
\providecommand \bibfield  [0]{\@secondoftwo}%
\providecommand \translation [1]{[#1]}%
\providecommand \BibitemOpen [0]{}%
\providecommand \bibitemStop [0]{}%
\providecommand \bibitemNoStop [0]{.\EOS\space}%
\providecommand \EOS [0]{\spacefactor3000\relax}%
\providecommand \BibitemShut  [1]{\csname bibitem#1\endcsname}%
\let\auto@bib@innerbib\@empty
\end{thebibliography}%


\begin{thebibliography}{99}

\bibitem[Bonderson et~al.(2008)Bonderson, Freedman, and
  Nayak]{PhysRevLett.101.010501}
Bonderson, P.; Freedman, M.; Nayak, C.
\newblock Measurement-Only Topological Quantum Computation.
\newblock {\em Phys. Rev. Lett.} {\bf 2008}, {\em 101},~010501.
\newblock {\url{https://doi.org/10.1103/PhysRevLett.101.010501}}.


\bibitem[Leung(2004)]{leung2004quantum}
Leung, D.W.
\newblock Quantum computation by measurements.
\newblock {\em Int. J. Quantum Inf.} {\bf 2004}, {\em 2},~33--43.

\bibitem[Skinner et~al.(2019)Skinner, Ruhman, and Nahum]{PhysRevX.9.031009}
Skinner, B.; Ruhman, J.; Nahum, A.
\newblock Measurement-Induced Phase Transitions in the Dynamics of
  Entanglement.
\newblock {\em Phys. Rev. X} {\bf 2019}, {\em 9},~031009.
\newblock {\url{https://doi.org/10.1103/PhysRevX.9.031009}}.

\bibitem[Li et~al.(2018)Li, Chen, and Fisher]{PhysRevB.98.205136}
Li, Y.; Chen, X.; Fisher, M.P.A.
\newblock Quantum Zeno effect and the many-body entanglement transition.
\newblock {\em Phys. Rev. B} {\bf 2018}, {\em 98},~205136.
\newblock {\url{https://doi.org/10.1103/PhysRevB.98.205136}}.

\bibitem[Li et~al.(2019)Li, Chen, and Fisher]{PhysRevB.100.134306}
Li, Y.; Chen, X.; Fisher, M.P.A.
\newblock Measurement-driven entanglement transition in hybrid quantum
  circuits.
\newblock {\em Phys. Rev. B} {\bf 2019}, {\em 100},~134306.
\newblock {\url{https://doi.org/10.1103/PhysRevB.100.134306}}.

\bibitem[Christopoulos et~al.(2023)Christopoulos, Le~Doussal, Bernard, and
  De~Luca]{PhysRevX.13.011043}
Christopoulos, A.; Le~Doussal, P.; Bernard, D.; De~Luca, A.
\newblock Universal Out-of-Equilibrium Dynamics of 1D Critical Quantum Systems
  Perturbed by Noise Coupled to Energy.
\newblock {\em Phys. Rev. X} {\bf 2023}, {\em 13},~011043.
\newblock {\url{https://doi.org/10.1103/PhysRevX.13.011043}}.

\bibitem[Hruza and Bernard(2023)]{PhysRevX.13.011045}
Hruza, L.; Bernard, D.
\newblock Coherent Fluctuations in Noisy Mesoscopic Systems, the Open Quantum
  SSEP, and Free Probability.
\newblock {\em Phys. Rev. X} {\bf 2023}, {\em 13},~011045.
\newblock {\url{https://doi.org/10.1103/PhysRevX.13.011045}}.

\bibitem[Bernard(2021)]{Bernard_2021}
Bernard, D.
\newblock Can the macroscopic fluctuation theory be quantized?
\newblock {\em J. Phys. A Math. Theor.} {\bf 2021}, {\em 54},~433001.
\newblock {\url{https://doi.org/10.1088/1751-8121/ac2597}}.

\bibitem[Zhou and Nahum(2019)]{PhysRevB.99.174205}
Zhou, T.; Nahum, A.
\newblock Emergent statistical mechanics of entanglement in random unitary
  circuits.
\newblock {\em Phys. Rev. B} {\bf 2019}, {\em 99},~174205.
\newblock {\url{https://doi.org/10.1103/PhysRevB.99.174205}}.

\bibitem[Gullans and Huse(2020)]{PhysRevLett.125.070606}
Gullans, M.J.; Huse, D.A.
\newblock Scalable Probes of Measurement-Induced Criticality.
\newblock {\em Phys. Rev. Lett.} {\bf 2020}, {\em 125},~070606.
\newblock {\url{https://doi.org/10.1103/PhysRevLett.125.070606}}.

\bibitem[D'Alessio et~al.(2016)D'Alessio, Kafri, Polkovnikov, and
  Rigol]{dalessio2016}
D'Alessio, L.; Kafri, Y.; Polkovnikov, A.; Rigol, M.
\newblock From quantum chaos and eigenstate thermalization to statistical
  mechanics and thermodynamics.
\newblock {\em Adv. Phys.} {\bf 2016}, {\em 65},~239--362.
\newblock {\url{https://doi.org/10.1080/00018732.2016.1198134}}.

\bibitem[Nahum et~al.(2017)Nahum, Ruhman, Vijay, and Haah]{PhysRevX.7.031016}
Nahum, A.; Ruhman, J.; Vijay, S.; Haah, J.
\newblock Quantum Entanglement Growth under Random Unitary Dynamics.
\newblock {\em Phys. Rev. X} {\bf 2017}, {\em 7},~031016.
\newblock {\url{https://doi.org/10.1103/PhysRevX.7.031016}}.

\bibitem[Kim and Huse(2013)]{PhysRevLett.111.127205}
Kim, H.; Huse, D.A.
\newblock Ballistic Spreading of Entanglement in a Diffusive Nonintegrable
  System.
\newblock {\em Phys. Rev. Lett.} {\bf 2013}, {\em 111},~127205.
\newblock {\url{https://doi.org/10.1103/PhysRevLett.111.127205}}.

\bibitem[Bardarson et~al.(2012)Bardarson, Pollmann, and
  Moore]{PhysRevLett.109.017202}
Bardarson, J.H.; Pollmann, F.; Moore, J.E.
\newblock Unbounded Growth of Entanglement in Models of Many-Body Localization.
\newblock {\em Phys. Rev. Lett.} {\bf 2012}, {\em 109},~017202.
\newblock {\url{https://doi.org/10.1103/PhysRevLett.109.017202}}.

\bibitem[Abanin et~al.(2019)Abanin, Altman, Bloch, and
  Serbyn]{RevModPhys.91.021001}
Abanin, D.A.; Altman, E.; Bloch, I.; Serbyn, M.
\newblock Colloquium: Many-body localization, thermalization, and entanglement.
\newblock {\em Rev. Mod. Phys.} {\bf 2019}, {\em 91},~021001.
\newblock {\url{https://doi.org/10.1103/RevModPhys.91.021001}}.

\bibitem[Calabrese and Cardy(2005)]{Calabrese_2005}
Calabrese, P.; Cardy, J.
\newblock Evolution of entanglement entropy in one-dimensional systems.
\newblock {\em J. Stat. Mech. Theory Exp.} {\bf 2005}, {\em 2005},~P04010.
\newblock {\url{https://doi.org/10.1088/1742-5468/2005/04/P04010}}.

\bibitem[Chan et~al.(2018)Chan, De~Luca, and Chalker]{PhysRevX.8.041019}
Chan, A.; De~Luca, A.; Chalker, J.T.
\newblock Solution of a Minimal Model for Many-Body Quantum Chaos.
\newblock {\em Phys. Rev. X} {\bf 2018}, {\em 8},~041019.
\newblock {\url{https://doi.org/10.1103/PhysRevX.8.041019}}.

\bibitem[Bertini et~al.(2019)Bertini, Kos, and Prosen]{PhysRevX.9.021033}
Bertini, B.; Kos, P.; Prosen, T.c.v.
\newblock Entanglement Spreading in a Minimal Model of Maximal Many-Body
  Quantum Chaos.
\newblock {\em Phys. Rev. X} {\bf 2019}, {\em 9},~021033.
\newblock {\url{https://doi.org/10.1103/PhysRevX.9.021033}}.

\bibitem[Fisher et~al.(2023)Fisher, Khemani, Nahum, and
  Vijay]{doi:10.1146/annurev-conmatphys-031720-030658}
Fisher, M.P.; Khemani, V.; Nahum, A.; Vijay, S.
\newblock Random Quantum Circuits.
\newblock {\em Annu. Rev. Condens. Matter Phys.} {\bf 2023}, {\em
  14},~335--379.
\newblock {\url{https://doi.org/10.1146/annurev-conmatphys-031720-030658}}.

\bibitem[Lunt et~al.(2021)Lunt, Szyniszewski, and Pal]{PhysRevB.104.155111}
Lunt, O.; Szyniszewski, M.; Pal, A.
\newblock Measurement-induced criticality and entanglement clusters: A study of
  one-dimensional and two-dimensional Clifford circuits.
\newblock {\em Phys. Rev. B} {\bf 2021}, {\em 104},~155111.
\newblock {\url{https://doi.org/10.1103/PhysRevB.104.155111}}.

\bibitem[Jian et~al.(2020)Jian, You, Vasseur, and Ludwig]{PhysRevB.101.104302}
Jian, C.M.; You, Y.Z.; Vasseur, R.; Ludwig, A.W.W.
\newblock Measurement-induced criticality in random quantum circuits.
\newblock {\em Phys. Rev. B} {\bf 2020}, {\em 101},~104302.
\newblock {\url{https://doi.org/10.1103/PhysRevB.101.104302}}.

\bibitem[Zabalo et~al.(2022)Zabalo, Gullans, Wilson, Vasseur, Ludwig,
  Gopalakrishnan, Huse, and Pixley]{PhysRevLett.128.050602}
Zabalo, A.; Gullans, M.J.; Wilson, J.H.; Vasseur, R.; Ludwig, A.W.W.;
  Gopalakrishnan, S.; Huse, D.A.; Pixley, J.H.
\newblock Operator Scaling Dimensions and Multifractality at
  Measurement-Induced Transitions.
\newblock {\em Phys. Rev. Lett.} {\bf 2022}, {\em 128},~050602.
\newblock {\url{https://doi.org/10.1103/PhysRevLett.128.050602}}.

\bibitem[Nahum and Wiese(2023)]{nahum2023renormalization}
Nahum, A.; Wiese, K.J.
\newblock Renormalization group for measurement and entanglement phase transitions.  
\newblock \emph{Phys. Rev. B} {\bf 2023}, {\em 108(10)},~104203. 
\newblock \url{https://journals.aps.org/prb/abstract/10.1103/PhysRevB.108.104203}.

\bibitem[Minato et~al.(2022)Minato, Sugimoto, Kuwahara, and
  Saito]{PhysRevLett.128.010603}
Minato, T.; Sugimoto, K.; Kuwahara, T.; Saito, K.
\newblock Fate of Measurement-Induced Phase Transition in Long-Range
  Interactions.
\newblock {\em Phys. Rev. Lett.} {\bf 2022}, {\em 128},~010603.
\newblock {\url{https://doi.org/10.1103/PhysRevLett.128.010603}}.

\bibitem[Zhou(2023)]{10.21468/SciPostPhysCore.6.1.023}
Zhou, Y.N.
\newblock {Generalized Lindblad master equation for measurement-induced phase
  transition}.
\newblock {\em SciPost Phys. Core} {\bf 2023}, {\em 6},~023.
\newblock {\url{https://doi.org/10.21468/SciPostPhysCore.6.1.023}}.

\bibitem[Tang and Zhu(2020)]{PhysRevResearch.2.013022}
Tang, Q.; Zhu, W.
\newblock Measurement-induced phase transition: A case study in the
  nonintegrable model by density-matrix renormalization group calculations.
\newblock {\em Phys. Rev. Res.} {\bf 2020}, {\em 2},~013022.
\newblock {\url{https://doi.org/10.1103/PhysRevResearch.2.013022}}.

\bibitem[Willsher et~al.(2022)Willsher, Liu, Moessner, and
  Knolle]{PhysRevB.106.024305}
Willsher, J.; Liu, S.W.; Moessner, R.; Knolle, J.
\newblock Measurement-induced phase transition in a chaotic classical many-body
  system.
\newblock {\em Phys. Rev. B} {\bf 2022}, {\em 106},~024305.
\newblock {\url{https://doi.org/10.1103/PhysRevB.106.024305}}.

\bibitem[Li et~al.(2023)Li, Zou, Glorioso, Altman, and
  Fisher]{PhysRevLett.130.220404}
Li, Y.; Zou, Y.; Glorioso, P.; Altman, E.; Fisher, M.P.A.
\newblock Cross Entropy Benchmark for Measurement-Induced Phase Transitions.
\newblock {\em Phys. Rev. Lett.} {\bf 2023}, {\em 130},~220404.
\newblock {\url{https://doi.org/10.1103/PhysRevLett.130.220404}}.

\bibitem[Yang et~al.(2022)Yang, Li, Fisher, and Chen]{PhysRevB.105.104306}
Yang, Z.C.; Li, Y.; Fisher, M.P.A.; Chen, X.
\newblock Entanglement phase transitions in random stabilizer tensor networks.
\newblock {\em Phys. Rev. B} {\bf 2022}, {\em 105},~104306.
\newblock {\url{https://doi.org/10.1103/PhysRevB.105.104306}}.

\bibitem[Choi et~al.(2020)Choi, Bao, Qi, and Altman]{PhysRevLett.125.030505}
Choi, S.; Bao, Y.; Qi, X.L.; Altman, E.
\newblock Quantum Error Correction in Scrambling Dynamics and
  Measurement-Induced Phase Transition.
\newblock {\em Phys. Rev. Lett.} {\bf 2020}, {\em 125},~030505.
\newblock {\url{https://doi.org/10.1103/PhysRevLett.125.030505}}.

\bibitem[Gullans and Huse(2020)]{PhysRevX.10.041020}
Gullans, M.J.; Huse, D.A.
\newblock Dynamical Purification Phase Transition Induced by Quantum
  Measurements.
\newblock {\em Phys. Rev. X} {\bf 2020}, {\em 10},~041020.
\newblock {\url{https://doi.org/10.1103/PhysRevX.10.041020}}.

\bibitem[Ticozzi and Viola(2014)]{Ticozzi2014}
Ticozzi, F.; Viola, L.
\newblock Quantum resources for purification and cooling: Fundamental limits
  and opportunities.
\newblock {\em Sci. Rep.} {\bf 2014}, {\em 4},~5192.
\newblock {\url{https://doi.org/10.1038/srep05192}}.

\bibitem[L\'oio et~al.(2023)L\'oio, De~Luca, De~Nardis, and
  Turkeshi]{loio2023purification}
L\'oio, H.; De~Luca, A.; De~Nardis, J.; Turkeshi, X.
\newblock Purification timescales in monitored fermions.
\newblock {\em Phys. Rev. B} {\bf 2023}, {\em 108},~L020306.
\newblock {\url{https://doi.org/10.1103/PhysRevB.108.L020306}}.

\bibitem[Kelly et~al.(2023)Kelly, Poschinger, Schmidt-Kaler, Fisher, and
  Marino]{kelly2023coherence}
Kelly, S.P.; Poschinger, U.; Schmidt-Kaler, F.; Fisher, M.; Marino, J.
\newblock Coherence requirements for quantum communication from hybrid circuit
  dynamics.
\newblock {\em SciPost Phys.} {\bf 2023}, {\em 15},~250.
\newblock {\url{https://doi.org/10.21468/SciPostPhys.15.6.250}}.

\bibitem[Vidal(2003)]{PhysRevLett.91.147902}
Vidal, G.
\newblock Efficient Classical Simulation of Slightly Entangled Quantum
  Computations.
\newblock {\em Phys. Rev. Lett.} {\bf 2003}, {\em 91},~147902.
\newblock {\url{https://doi.org/10.1103/PhysRevLett.91.147902}}.

\bibitem[Vidal(2004)]{PhysRevLett.93.040502}
Vidal, G.
\newblock Efficient Simulation of One-Dimensional Quantum Many-Body Systems.
\newblock {\em Phys. Rev. Lett.} {\bf 2004}, {\em 93},~040502.
\newblock {\url{https://doi.org/10.1103/PhysRevLett.93.040502}}.

\bibitem[Claeys et~al.(2022)Claeys, Henry, Vicary, and
  Lamacraft]{PhysRevResearch.4.043212}
Claeys, P.W.; Henry, M.; Vicary, J.; Lamacraft, A.
\newblock Exact dynamics in dual-unitary quantum circuits with projective
  measurements.
\newblock {\em Phys. Rev. Res.} {\bf 2022}, {\em 4},~043212.
\newblock {\url{https://doi.org/10.1103/PhysRevResearch.4.043212}}.

\bibitem[Nahum and Skinner(2020)]{PhysRevResearch.2.023288}
Nahum, A.; Skinner, B.
\newblock Entanglement and dynamics of diffusion-annihilation processes with
  Majorana defects.
\newblock {\em Phys. Rev. Res.} {\bf 2020}, {\em 2},~023288.
\newblock {\url{https://doi.org/10.1103/PhysRevResearch.2.023288}}.

\bibitem[Cao et~al.(2019)Cao, Tilloy, and Luca]{10.21468/SciPostPhys.7.2.024}
Cao, X.; Tilloy, A.; Luca, A.D.
\newblock {Entanglement in a fermion chain under continuous monitoring}.
\newblock {\em SciPost Phys.} {\bf 2019}, {\em 7},~024.
\newblock {\url{https://doi.org/10.21468/SciPostPhys.7.2.024}}.

\bibitem[Fidkowski et~al.(2021)Fidkowski, Haah, and
  Hastings]{Fidkowski2021howdynamicalquantum}
Fidkowski, L.; Haah, J.; Hastings, M.B.
\newblock How {D}ynamical {Q}uantum {M}emories {F}orget.
\newblock {\em {Quantum}} {\bf 2021}, {\em 5},~382.
\newblock {\url{https://doi.org/10.22331/q-2021-01-17-382}}.

\bibitem[Coppola et~al.(2022)Coppola, Tirrito, Karevski, and
  Collura]{PhysRevB.105.094303}
Coppola, M.; Tirrito, E.; Karevski, D.; Collura, M.
\newblock Growth of entanglement entropy under local projective measurements.
\newblock {\em Phys. Rev. B} {\bf 2022}, {\em 105},~094303.
\newblock {\url{https://doi.org/10.1103/PhysRevB.105.094303}}.

\bibitem[Santini et~al.(2023)Santini, Solfanelli, Gherardini, and
  Giachetti]{Santini2023Observation}
Santini, A.; Solfanelli, A.; Gherardini, S.; Giachetti, G.
\newblock Observation of partial and infinite-temperature thermalization
  induced by repeated measurements on a quantum hardware.
\newblock {\em J. Phys. Commun.} {\bf 2023}, {\em 7},~065007.
\newblock {\url{https://doi.org/10.1088/2399-6528/acdd4f}}.

\bibitem[Piccitto et~al.(2022)Piccitto, Russomanno, and
  Rossini]{PhysRevB.105.064305}
Piccitto, G.; Russomanno, A.; Rossini, D.
\newblock Entanglement transitions in the quantum Ising chain: A comparison
  between different unravelings of the same Lindbladian.
\newblock {\em Phys. Rev. B} {\bf 2022}, {\em 105},~064305.
\newblock {\url{https://doi.org/10.1103/PhysRevB.105.064305}}.

\bibitem[Turkeshi et~al.(2022)Turkeshi, Dalmonte, Fazio, and
  Schir\`o]{PhysRevB.105.L241114}
Turkeshi, X.; Dalmonte, M.; Fazio, R.; Schir\`o, M.
\newblock Entanglement transitions from stochastic resetting of non-Hermitian
  quasiparticles.
\newblock {\em Phys. Rev. B} {\bf 2022}, {\em 105},~L241114.
\newblock {\url{https://doi.org/10.1103/PhysRevB.105.L241114}}.

\bibitem[Turkeshi et~al.(2021)Turkeshi, Biella, Fazio, Dalmonte, and
  Schir\'o]{PhysRevB.103.224210}
Turkeshi, X.; Biella, A.; Fazio, R.; Dalmonte, M.; Schir\'o, M.
\newblock Measurement-induced entanglement transitions in the quantum Ising
  chain: From infinite to zero clicks.
\newblock {\em Phys. Rev. B} {\bf 2021}, {\em 103},~224210.
\newblock {\url{https://doi.org/10.1103/PhysRevB.103.224210}}.

\bibitem[Alberton et~al.(2021)Alberton, Buchhold, and
  Diehl]{PhysRevLett.126.170602}
Alberton, O.; Buchhold, M.; Diehl, S.
\newblock Entanglement Transition in a Monitored Free--Fermion Chain: From
  Extended Criticality to Area Law.
\newblock {\em Phys. Rev. Lett.} {\bf 2021}, {\em 126},~170602.
\newblock {\url{https://doi.org/10.1103/PhysRevLett.126.170602}}.

\bibitem[Buchhold et~al.(2021)Buchhold, Minoguchi, Altland, and
  Diehl]{PhysRevX.11.041004}
Buchhold, M.; Minoguchi, Y.; Altland, A.; Diehl, S.
\newblock Effective Theory for the Measurement-Induced Phase Transition of
  Dirac Fermions.
\newblock {\em Phys. Rev. X} {\bf 2021}, {\em 11},~041004.
\newblock {\url{https://doi.org/10.1103/PhysRevX.11.041004}}.

\bibitem[M\"uller et~al.(2022)M\"uller, Diehl, and
  Buchhold]{PhysRevLett.128.010605}
M\"uller, T.; Diehl, S.; Buchhold, M.
\newblock Measurement-Induced Dark State Phase Transitions in Long-Ranged
  Fermion Systems.
\newblock {\em Phys. Rev. Lett.} {\bf 2022}, {\em 128},~010605.
\newblock {\url{https://doi.org/10.1103/PhysRevLett.128.010605}}.

\bibitem[Ladewig et~al.(2022)Ladewig, Diehl, and
  Buchhold]{PhysRevResearch.4.033001}
Ladewig, B.; Diehl, S.; Buchhold, M.
\newblock Monitored open fermion dynamics: Exploring the interplay of
  measurement, decoherence, and free Hamiltonian evolution.
\newblock {\em Phys. Rev. Res.} {\bf 2022}, {\em 4},~033001.
\newblock {\url{https://doi.org/10.1103/PhysRevResearch.4.033001}}.

\bibitem[Lucas et~al.(2023)Lucas, Piroli, De~Nardis, and
  De~Luca]{PhysRevA.107.032215}
Lucas, M.; Piroli, L.; De~Nardis, J.; De~Luca, A.
\newblock Generalized deep thermalization for free fermions.
\newblock {\em Phys. Rev. A} {\bf 2023}, {\em 107},~032215.
\newblock {\url{https://doi.org/10.1103/PhysRevA.107.032215}}.

\bibitem[Fava et~al.(2023)Fava, Piroli, Swann, Bernard, and
  Nahum]{fava2023nonlinear}
Fava, M.; Piroli, L.; Swann, T.; Bernard, D.; Nahum, A.
\newblock Nonlinear Sigma Models for Monitored Dynamics of Free Fermions.
\newblock {\em Phys. Rev. X} {\bf 2023}, {\em 13},~041045.
\newblock {\url{https://doi.org/10.1103/PhysRevX.13.041045}}.

\bibitem[Poboiko et~al.(2023)Poboiko, P\"opperl, Gornyi, and
  Mirlin]{poboiko2023theory}
Poboiko, I.; P\"opperl, P.; Gornyi, I.V.; Mirlin, A.D.
\newblock Theory of Free Fermions under Random Projective Measurements.
\newblock {\em Phys. Rev. X} {\bf 2023}, {\em 13},~041046.
\newblock {\url{https://doi.org/10.1103/PhysRevX.13.041046}}.

\bibitem[M{\'e}zard et~al.(1987)M{\'e}zard, Parisi, and
  Virasoro]{mezard1987spin}
M{\'e}zard, M.; Parisi, G.; Virasoro, M.A.
\newblock {\em Spin Glass Theory and Beyond}; World Scientific
  Publishing Company:  Singapore,  1987; Volume~9.

\bibitem[Bray and Moore(1980)]{A_J_Bray_1980}
Bray, A.J.; Moore, M.A.
\newblock Replica theory of quantum spin glasses.
\newblock {\em J. Phys. C Solid State Phys.} {\bf 1980}, {\em 13},~L655.
\newblock {\url{https://doi.org/10.1088/0022-3719/13/24/005}}.

\bibitem[Lopez-Piqueres et~al.(2020)Lopez-Piqueres, Ware, and
  Vasseur]{PhysRevB.102.064202}
Lopez-Piqueres, J.; Ware, B.; Vasseur, R.
\newblock Mean-field entanglement transitions in random tree tensor networks.
\newblock {\em Phys. Rev. B} {\bf 2020}, {\em 102},~064202.
\newblock {\url{https://doi.org/10.1103/PhysRevB.102.064202}}.

\bibitem[Vasseur et~al.(2019)Vasseur, Potter, You, and
  Ludwig]{PhysRevB.100.134203}
Vasseur, R.; Potter, A.C.; You, Y.Z.; Ludwig, A.W.W.
\newblock Entanglement transitions from holographic random tensor networks.
\newblock {\em Phys. Rev. B} {\bf 2019}, {\em 100},~134203.
\newblock {\url{https://doi.org/10.1103/PhysRevB.100.134203}}.

\bibitem[Bentsen et~al.(2021)Bentsen, Sahu, and
  Swingle]{bentsen2021measurement}
Bentsen, G.S.; Sahu, S.; Swingle, B.
\newblock Measurement-induced purification in large-N hybrid Brownian circuits.
\newblock {\em Phys. Rev. B} {\bf 2021}, {\em 104},~094304.
\newblock {\url{https://doi.org/10.1103/PhysRevB.104.094304}}.

\bibitem[Jian et~al.(2021)Jian, Liu, Chen, Swingle, and
  Zhang]{PhysRevLett.127.140601}
Jian, S.K.; Liu, C.; Chen, X.; Swingle, B.; Zhang, P.
\newblock Measurement-Induced Phase Transition in the Monitored
  Sachdev-Ye-Kitaev Model.
\newblock {\em Phys. Rev. Lett.} {\bf 2021}, {\em 127},~140601.
\newblock {\url{https://doi.org/10.1103/PhysRevLett.127.140601}}.

\bibitem[Nahum et~al.(2021)Nahum, Roy, Skinner, and
  Ruhman]{PRXQuantum.2.010352}
Nahum, A.; Roy, S.; Skinner, B.; Ruhman, J.
\newblock Measurement and Entanglement Phase Transitions in All-To-All Quantum
  Circuits, on Quantum Trees, and in Landau-Ginsburg Theory.
\newblock {\em PRX Quantum} {\bf 2021}, {\em 2},~010352.
\newblock {\url{https://doi.org/10.1103/PRXQuantum.2.010352}}.

\bibitem[Fisher(1937)]{fisher1937wave}
Fisher, R.A.
\newblock The wave of advance of advantageous genes.
\newblock {\em Ann. Eugen.} {\bf 1937}, {\em 7},~355--369.
\newblock {\url{https://doi.org/10.1111/j.1469-1809.1937.tb02153.x}}.


\bibitem[Kolmogorov et~al.(1937)]{KPP}
Kolmogorov, A.N.; Petrovskii, I.; Piskunov, N.S.
\newblock A Study of the Diffusion Equation with Increase in the Amount of Substance, and its Application to a Biological Problem.
\newblock  \emph{Mosc. Univ. Math. Bull.} {\bf 1991}, {\em 6(1)},~1--25. 
\newblock {\url{https://cir.nii.ac.jp/crid/1573668924865104768}}.

\bibitem[Derrida and Spohn(1988)]{Derrida1988}
Derrida, B.; Spohn, H.
\newblock Polymers on disordered trees, spin glasses, and traveling waves.
\newblock {\em J. Stat. Phys.} {\bf 1988}, {\em 51},~817--840.
\newblock {\url{https://doi.org/10.1007/BF01014886}}.

\bibitem[Giachetti and De~Luca(2023)]{giachetti2023elusive}
Giachetti, G.; De~Luca, A.
\newblock Elusive phase transition in the replica limit of monitored systems. \emph{arXiv}
  \textbf{2023}, arXiv:2306.12166.  
\newblock {\url{http://arxiv.org/abs/2306.12166}}.

\bibitem[Schomerus(2022)]{schomerus2022noisy}
Schomerus, H.
\newblock Noisy monitored quantum dynamics of ergodic multi-qubit systems.
\newblock {\em J. Phys. A Math. Theor.} {\bf 2022}, {\em 55},~214001.
\newblock {\url{https://doi.org/10.1088/1751-8121/ac6320}}.

\bibitem[Caves and Milburn(1987)]{PhysRevA.36.5543}
Caves, C.M.; Milburn, G.J.
\newblock Quantum-mechanical model for continuous position measurements.
\newblock {\em Phys. Rev. A} {\bf 1987}, {\em 36},~5543--5555.
\newblock {\url{https://doi.org/10.1103/PhysRevA.36.5543}}.

\bibitem[Nielsen and Chuang(2002)]{nielsen2002quantum}
Nielsen, M.A.; Chuang, I.
\newblock \emph{Quantum Computation and Quantum Information}; Cambridge University Press: Cambridge, UK,  2002.

\bibitem[Choi(1975)]{CHOI1975285}
Choi, M.D.
\newblock Completely positive linear maps on complex matrices.
\newblock {\em Linear Algebra Its Appl.} {\bf 1975}, {\em 10},~285--290.
\newblock {\url{https://doi.org/10.1016/0024-3795(75)90075-0}}.

\bibitem[Kraus(1971)]{KRAUS1971311}
Kraus, K.
\newblock General state changes in quantum theory.
\newblock {\em Ann. Phys.} {\bf 1971}, {\em 64},~311--335.
\newblock {\url{https://doi.org/10.1016/0003-4916(71)90108-4}}.

\bibitem[Di\'osi et~al.(1998)Di\'osi, Gisin, and Strunz]{PhysRevA.58.1699}
Di\'osi, L.; Gisin, N.; Strunz, W.T.
\newblock Non-Markovian quantum state diffusion.
\newblock {\em Phys. Rev. A} {\bf 1998}, {\em 58},~1699--1712.
\newblock {\url{https://doi.org/10.1103/PhysRevA.58.1699}}.

\bibitem[Gisin and Percival(1992)]{Gisin_1992}
Gisin, N.; Percival, I.C.
\newblock The quantum-state diffusion model applied to open systems.
\newblock {\em J. Phys. A Math. Gen.} {\bf 1992}, {\em 25},~5677.
\newblock {\url{https://doi.org/10.1088/0305-4470/25/21/023}}.

\bibitem[Dyson(2004)]{10.1063/1.1703862}
Dyson, F.J.
\newblock {A Brownian‐Motion Model for the Eigenvalues of a Random Matrix}.
\newblock {\em J. Math. Phys.} {\bf 2004}, {\em 3},~1191--1198.
\newblock {\url{https://doi.org/10.1063/1.1703862}}.

\bibitem[Gauti{\'e} et~al.(2021)Gauti{\'e}, Bouchaud, and
  Le~Doussal]{gautie2021matrix}
Gauti{\'e}, T.; Bouchaud, J.P.; Le~Doussal, P.
\newblock Matrix Kesten recursion, inverse-Wishart ensemble and fermions in a
  Morse potential.
\newblock {\em J. Phys. A Math. Theor.} {\bf 2021}, {\em 54},~255201.
\newblock {\url{https://doi.org/10.1088/1751-8121/abfc7f}}.

\bibitem[Karlin and McGregor(1959)]{pjm/1103038889}
Karlin, S.; McGregor, J.
\newblock {Coincidence probabilities.}
\newblock {\em Pac. J. Math.} {\bf 1959}, {\em 9},~1141 -- 1164.

\bibitem[Smith et~al.(2021)Smith, Doussal, Majumdar, and
  Schehr]{10.21468/SciPostPhys.11.6.110}
Smith, N.R.; Doussal, P.L.; Majumdar, S.N.; Schehr, G.
\newblock {Full counting statistics for interacting trapped fermions}.
\newblock {\em SciPost Phys.} {\bf 2021}, {\em 11},~110.
\newblock {\url{https://doi.org/10.21468/SciPostPhys.11.6.110}}.

\bibitem[Ipsen and Schomerus(2016)]{Ipsen_2016}
Ipsen, J.R.; Schomerus, H.
\newblock Isotropic Brownian motions over complex fields as a solvable model
  for May–Wigner stability analysis.
\newblock {\em J. Phys. A Math. Theor.} {\bf 2016}, {\em 49},~385201.
\newblock {\url{https://doi.org/10.1088/1751-8113/49/38/385201}}.

\bibitem[Mergny and Majumdar(2021)]{Mergny_2021}
Mergny, P.; Majumdar, S.N.
\newblock Stability of large complex systems with heterogeneous relaxation
  dynamics.
\newblock {\em J. Stat. Mech. Theory Exp.} {\bf 2021}, {\em 2021},~123301.
\newblock {\url{https://doi.org/10.1088/1742-5468/ac3b47}}.

\bibitem[Derrida and Hilhorst(1983)]{derrida_singular_1983}
Derrida, B.; Hilhorst, H.J.
\newblock Singular Behaviour of Certain Infinite Products of Random 2
  {\texttimes} 2 Matrices.
\newblock {\em J. Phys. A Math. Gen.} {\bf 1983}, {\em 16},~2641--2654.
\newblock {\url{https://doi.org/10.1088/0305-4470/16/12/013}}.

\bibitem[Bouchard et~al.(1986)Bouchard, Georges, Hansel, Doussal, and
  Maillard]{bouchard_rigorous_1986}
Bouchard, J.P.; Georges, A.; Hansel, D.; Doussal, P.L.; Maillard, J.M.
\newblock Rigorous Bounds and the Replica Method for Products of Random
  Matrices.
\newblock {\em J. Phys. A Math. Gen.} {\bf 1986}, {\em 19},~L1145--L1152.
\newblock {\url{https://doi.org/10.1088/0305-4470/19/18/006}}.

\bibitem[Ipsen(2015)]{ipsen_products_2015}
Ipsen, J.R.
\newblock Products of {{Independent Gaussian Random Matrices}}. \emph{arXiv}  \textbf{2015}, arXiv:1510.06128.
\newblock {\url{https://doi.org/10.48550/arXiv.1510.06128}}.

\bibitem[Haake(1991)]{haake1991quantum}
Haake, F.
\newblock {\em Quantum Signatures of Chaos}; Springer:  Berlin/Heidelberg, Germany,  1991.

\bibitem[Forrester(2020)]{forrester2020global}
Forrester, P.J.
\newblock Global and local scaling limits for the $\beta = 2$ Stieltjes--Wigert
  random matrix ensemble. \emph{arXiv} \textbf{2020}, arXiv:math-ph/2011.11783.

\bibitem[Macdonald(1998)]{macdonald1998symmetric}
Macdonald, I.G.
\newblock {\em Symmetric Functions and Hall Polynomials}; Oxford University
  Press: Oxford, UK,  1998.

\bibitem[Forrester(2019)]{doi:10.1142/S2010326319300018}
Forrester, P.J.
\newblock Meet Andréief, Bordeaux 1886, and Andreev, Kharkov 1882–1883.
\newblock {\em Random Matrices Theory Appl.} {\bf 2019}, {\em
  8},~1930001.
\newblock {\url{https://doi.org/10.1142/S2010326319300018}}.

\bibitem[Jonnadula et~al.(2021)Jonnadula, Keating, and
  Mezzadri]{10.1063/5.0048364}
Jonnadula, B.; Keating, J.P.; Mezzadri, F.
\newblock {Symmetric function theory and unitary invariant ensembles}.
\newblock {\em J. Math. Phys.} {\bf 2021}, {\em 62},~093512.
\newblock {\url{https://doi.org/10.1063/5.0048364}}.

\bibitem[{Wikipedia contributors}(2023)]{enwiki:1177949416}
{Wikipedia Contributors}.
\newblock Hypergeometric Function---{Wikipedia}{,} The Free Encyclopedia. 2023.
\newblock
   Available online: \url{https://en.wikipedia.org/w/index.php?title=Hypergeometric_function&oldid=1177949416}  (accessed on 22 December 2023).

\bibitem[Brézin and Hikami(1996)]{BREZIN1996697}
Brézin, E.; Hikami, S.
\newblock Correlations of nearby levels induced by a random potential.
\newblock {\em Nucl. Phys. B} {\bf 1996}, {\em 479},~697--706.
\newblock {\url{https://doi.org/10.1016/0550-3213(96)00394-X}}.

\bibitem[Br\'ezin and Hikami(1998)]{PhysRevE.58.7176}
Br\'ezin, E.; Hikami, S.
\newblock Level spacing of random matrices in an external source.
\newblock {\em Phys. Rev. E} {\bf 1998}, {\em 58},~7176--7185.
\newblock {\url{https://doi.org/10.1103/PhysRevE.58.7176}}.

\bibitem[Johansson(2001)]{johansson2001universality}
Johansson, K.J.
\newblock Universality of the Local Spacing Distribution in Certain Ensembles
  of Hermitian Wigner Matrices.
\newblock {\em Commun. Math. Phys.} {\bf 2001}, {\em 215},~683--705.
\newblock {\url{https://doi.org/10.1007/s002200000328}}.

\bibitem[Krajenbrink et~al.(2021)Krajenbrink, Le~Doussal, and
  O'Connell]{krajenbrink2021tilted}
Krajenbrink, A.; Le~Doussal, P.; O'Connell, N.
\newblock Tilted elastic lines with columnar and point disorder, non-Hermitian
  quantum mechanics, and spiked random matrices: Pinning and localization.
\newblock {\em Phys. Rev. E} {\bf 2021}, {\em 103},~042120.
\newblock {\url{https://doi.org/10.1103/PhysRevE.103.042120}}.

\bibitem[Claeys et~al.(2018)Claeys, Kuijlaars, Liechty, and
  Wang]{claeys2018propagation}
Claeys, T.; Kuijlaars, A.B.; Liechty, K.; Wang, D.
\newblock Propagation of singular behavior for Gaussian perturbations of random
  matrices.
\newblock {\em Commun. Math. Phys.} {\bf 2018}, {\em 362},~1--54.
\newblock {\url{https://doi.org/10.1007/s00220-018-3195-8}}.

\bibitem[Claeys and Wang(2014)]{claeys2014random}
Claeys, T.; Wang, D.
\newblock Random matrices with equispaced external source.
\newblock {\em Commun. Math. Phys.} {\bf 2014}, {\em 328},~1023--1077.
\newblock {\url{https://doi.org/10.1007/s00220-014-1988-y}}.

\bibitem[Forrester(1994)]{Forrester1994}
Forrester, P.J.
\newblock Properties of an exact crystalline many-body ground state.
\newblock {\em J. Stat. Phys.} {\bf 1994}, {\em 76},~331--346.
\newblock {\url{https://doi.org/10.1007/BF02188665}}.

\bibitem[Akemann et~al.(2013)Akemann, Kieburg, and Wei]{Akemann_2013}
Akemann, G.; Kieburg, M.; Wei, L.
\newblock Singular value correlation functions for products of Wishart random
  matrices.
\newblock {\em J. Phys. A Math. Theor.} {\bf 2013}, {\em 46},~275205.
\newblock {\url{https://doi.org/10.1088/1751-8113/46/27/275205}}.

\bibitem[{De Luca} et~al.(2023){De Luca}, Liu, Nahum, and
  Zhou]{deluca2023universality}
{De Luca}, A.; Liu, C.; Nahum, A.; Zhou, T.
\newblock Universality classes for purification in nonunitary quantum
  processes. \emph{arXiv}  \textbf{2023}, arXiv:2312.17744.

\bibitem[Flack et~al.(2023)Flack, Le~Doussal, Majumdar, and
  Schehr]{flack2023out}
Flack, A.; Le~Doussal, P.; Majumdar, S.N.; Schehr, G.
\newblock Out-of-equilibrium dynamics of repulsive ranked diffusions: The
  expanding crystal.
\newblock {\em Phys. Rev. E} {\bf 2023}, {\em 107},~064105.
\newblock {\url{https://doi.org/10.1103/PhysRevE.107.064105}}.

\bibitem[Le~Doussal(2022)]{ledoussal2022ranked}
Le~Doussal, P.
\newblock Ranked diffusion, delta Bose gas, and Burgers equation.
\newblock {\em Phys. Rev. E} {\bf 2022}, {\em 105},~L012103.
\newblock {\url{https://doi.org/10.1103/PhysRevE.105.L012103}}.

\bibitem[Bulchandani et~al.(2024)Bulchandani, Sondhi, and
  Chalker]{bulchandani2024random}
Bulchandani, V.B.; Sondhi, S.L.; Chalker, J.T.
\newblock Random-matrix models of monitored quantum circuits. 
\newblock {\em J. Stat. Phys.} {\bf 2024}, {\em 191(5)},~55.
\newblock {\url{https://link.springer.com/article/10.1007/s10955-024-03273-0}}

\bibitem[Livan et~al.(2018)Livan, Novaes, and Vivo]{livan2018introduction}
Livan, G.; Novaes, M.; Vivo, P.
\newblock Introduction to random matrices theory and practice.
\newblock {\em Monogr. Award} {\bf 2018}, {\em 63},~54--57.

\bibitem[Potters and Bouchaud(2020)]{potters2020first}
Potters, M.; Bouchaud, J.P.
\newblock {\em A First Course in Random Matrix Theory: For Physicists,
  Engineers and Data Scientists}; Cambridge University Press: Cambridge, UK,  2020.

\bibitem[Liu et~al.(2023)Liu, Wang, and Wang]{Liu2023}
Liu, D.Z.; Wang, D.; Wang, Y.
\newblock Lyapunov Exponent, Universality and Phase Transition for Products of
  Random Matrices.
\newblock {\em Commun. Math. Phys.} {\bf 2023}, {\em 399},~1811--1855.
\newblock {\url{https://doi.org/10.1007/s00220-022-04584-7}}.

\bibitem[Akemann et~al.(2020)Akemann, Burda, and Kieburg]{PhysRevE.102.052134}
Akemann, G.; Burda, Z.; Kieburg, M.
\newblock Universality of local spectral statistics of products of random
  matrices.
\newblock {\em Phys. Rev. E} {\bf 2020}, {\em 102},~052134.
\newblock {\url{https://doi.org/10.1103/PhysRevE.102.052134}}.

\end{thebibliography}
\end{document}